\documentclass[onecolumn,superscriptaddress,floatfix,fleqn,aps,pra]{revtex4-2}

\usepackage{natbib}
\usepackage{bm}
\usepackage{hyperref}
\usepackage{graphicx}
\usepackage{bm}	
\usepackage{color}
\usepackage{xcolor}
\usepackage{epsfig}
\usepackage{amsmath}
\usepackage{amssymb}
\usepackage{longtable}
\usepackage{float}
\usepackage{mathtools}
\usepackage{xparse}
\usepackage{times}
\usepackage{braket}
\usepackage{xspace}

\usepackage{siunitx}

\newcommand{\MaxCut}{\textsc{Max}-\textsc{Cut}\xspace}
\DeclareMathOperator{\Tr}{Tr}
\newcommand{\imag}{\mathrm{i}}

\newcommand{\RR}{\mathbb{R}}
\DeclareMathOperator*{\EE}{\mathbb{E}}
\newcommand{\QAOA}{\textsc{qaoa}\xspace}

\begin{document}

\title{Expectation Values from the Single-Layer Quantum Approximate Optimization Algorithm on Ising Problems}
\author{Asier~Ozaeta}
\email{asier.ozaeta@qcware.com}
\affiliation{QC Ware Corp., 550 Hamilton Ave, Palo Alto, CA 94301, USA}

\author{Wim~van~Dam}
\email{wim.van.dam@qcware.com}
\affiliation{QC Ware Corp., 550 Hamilton Ave, Palo Alto, CA 94301, USA}
\affiliation{Department of Computer Science, University of California, Santa Barbara, CA 93106, USA}
\affiliation{Department of Physics, University of California, Santa Barbara, CA 93106, USA}

\author{Peter~L.~McMahon}
\email{pmcmahon@cornell.edu}
\affiliation{School of Applied and Engineering Physics, Cornell University, Ithaca, NY 14853, USA}
\affiliation{QC Ware Corp., 550 Hamilton Ave, Palo Alto, CA 94301, USA}

\date{\today}

\begin{abstract}

We report on the energy-expectation-value landscapes produced by the single-layer ($p=1$) Quantum Approximate Optimization Algorithm (\QAOA) when being used to solve Ising problems. 
The landscapes are obtained using an analytical formula that we derive. 
The formula allows us to predict the landscape for any given Ising problem instance and consequently predict the optimal \QAOA parameters for heuristically solving that instance using the single-layer \QAOA. 
We have validated our analytical formula by showing that it accurately reproduces the landscapes published in recent experimental reports. 
We then applied our methods to address the question: \emph{how well is the single-layer \QAOA able to solve large benchmark problem instances?} We used our analytical formula to calculate the optimal energy-expectation values for benchmark \MaxCut problems containing up to $\num{7000}$ vertices and $\num{41459}$ edges. 
We also calculated the optimal energy expectations for general Ising problems with up to $\num{100000}$ vertices and $\num{150000}$ edges. 
Our results provide an estimate for how well the single-layer \QAOA may work when run on a quantum computer with thousands of qubits. 
In addition to providing performance estimates when optimal angles are used, we are able to use our analytical results to investigate the difficulties one may encounter when running the \QAOA in practice for different classes of Ising instances. 
We find that depending on the parameters of the Ising Hamiltonian, the expectation-value landscapes can be rather complex, with sharp features that necessitate highly accurate rotation gates in order for the \QAOA to be run optimally on quantum hardware. 
We also present analytical results that explain some of the qualitative landscape features that are observed numerically.

\end{abstract}

\maketitle

\tableofcontents

\section{Introduction}

The Quantum Approximate Optimization Algorithm (\QAOA) \cite{farhi2014quantum} is a variational quantum algorithm \cite{moll2018quantum} that is able to find approximate solutions to combinatorial-optimization problems \cite{crooks2018performance}, and can also used to heuristically find optimal solutions \cite{zhou2018quantum}.

The basic structure of the \QAOA, and some precursor quantum-optimization algorithms \cite{hogg2000quantum,trugenberger2002quantum}, is the application of an alternating series of unitary operations. These unitary operations correspond to applications of the cost function (cost Hamiltonian) that one is attempting to optimize for, and of a \emph{mixer} or \emph{driver} Hamiltonian. The application of these unitaries is intended to evolve the state of the quantum computer such that it has substantial overlap with computational-basis states representing low-energy configurations of the variables being optimized over. In the \QAOA, each unitary is parameterized by a single real number (often referred to in the \QAOA literature as an \emph{angle}), and these parameters are chosen to optimize the quality of the solutions output by the algorithm. In order for \QAOA to be effective it suffices to find good enough parameters, not necessarily optimal. However, the effectiveness of the \QAOA will be directly related to the quality of these parameters.A crucial question is how the choice of the \QAOA parameters affects the solutions the \QAOA is able to find, and hence how to best pick the parameters to be used when running the \QAOA.

In this paper we will discuss the \QAOA in the setting of solving (classical) Ising optimization problems, i.e., where the optimization problem's cost function takes the form $\sum_{(i,j)\in E} J_{ij} s_i s_j + \sum_{i\in V} h_i s_i$, and $s_i \in \{-1,+1\}$. 
We note that \MaxCut problems are a subclass of Ising-optimization problems, for which the external-field terms $h_i$ are zero and the couplings $J_{ij}$ are $-1$. Our results are however applicable to the solution of any NP-hard problem by the \QAOA, via an appropriate mapping \cite{lucas2014ising}.

The \QAOA has recently been experimentally demonstrated with up to 23 qubits on Ising-optimization problems with no external-field terms \cite{google_qaoa}. 
The experimental results in \cite{google_qaoa} include optimization landscapes for the case where the \QAOA has only a single \emph{layer} (a single application of the cost-function Hamiltonian, followed by a single application of the mixer Hamiltonian). 
These landscapes are plots of the expected value of the cost-function Hamiltonian as a function of the two \QAOA parameters. 
In our work we present an analytical formula that is able to compute the energy landscape of the single-layer \QAOA for any Ising-problem instance. 
We compare our analytical results with the numerical results from \cite{google_qaoa} (which include both numerical results from classical simulations of the \QAOA, and measured data from running the \QAOA on a quantum processor), and find excellent agreement---showing that our analytical formula is able to predict the energy landscape. 
Our analytical formula gives us the ability to analytically optimize the \QAOA parameters---this enables one to compute the optimal \QAOA parameters classically, and then only run the \QAOA on a quantum processor with the optimal set of angles \cite{streif2020training}. 
This yields a large speedup versus the current standard mode of operating the single-layer \QAOA, where the parameters are optimized numerically via an iterative process or via performing a brute-force search of the landscape.

In addition to replicating the landscapes for the three problem instances given in \cite{google_qaoa}, we also explore the landscapes our analytical formula obtains for a range of other Ising problems, some of which we discover have qualitatively different features than the landscapes published in the \QAOA literature so far.

One of the key benefits of having an analytical formula is that we can study the performance of the \QAOA on problems using far more qubits than we can classically simulate or that we have experimental hardware containing. 
We use our analytical formula to predict the performance of the \QAOA on the G-set \MaxCut benchmark instances \cite{G_set}, which includes problems having up to $\num{7000}$ vertices (qubits), as well as on Ising instances with up to $\num{100000}$ vertices.
These results give us an indication of how well the \QAOA may perform on problem instances with sizes far beyond the $\sim 50$ spins that is the current limit for direct testing of the \QAOA.

There has been previous work on studying the \QAOA parameters and energy-expectation landscapes analytically in the single-layer setting: \cite{wang2018quantum} derives a formula for the energy expectation as a function of the \QAOA parameters for the specific case of \MaxCut on unweighted graphs, and \cite{Bravyi2019ObstaclesTS} generalized this formula for the energy expectation for the \QAOA to the case of weighted graphs. 
Our work builds on this as we include in our analysis the case of non-zero external-field terms in the Ising Hamiltonian. 
This generalization allows us to assess the energy expectation for the single-layer \QAOA solving any Ising problem. 

\section{Analytical Formulae for Expected Energies}

\begin{figure}[t]
\centering
\includegraphics[width=0.5\textwidth]{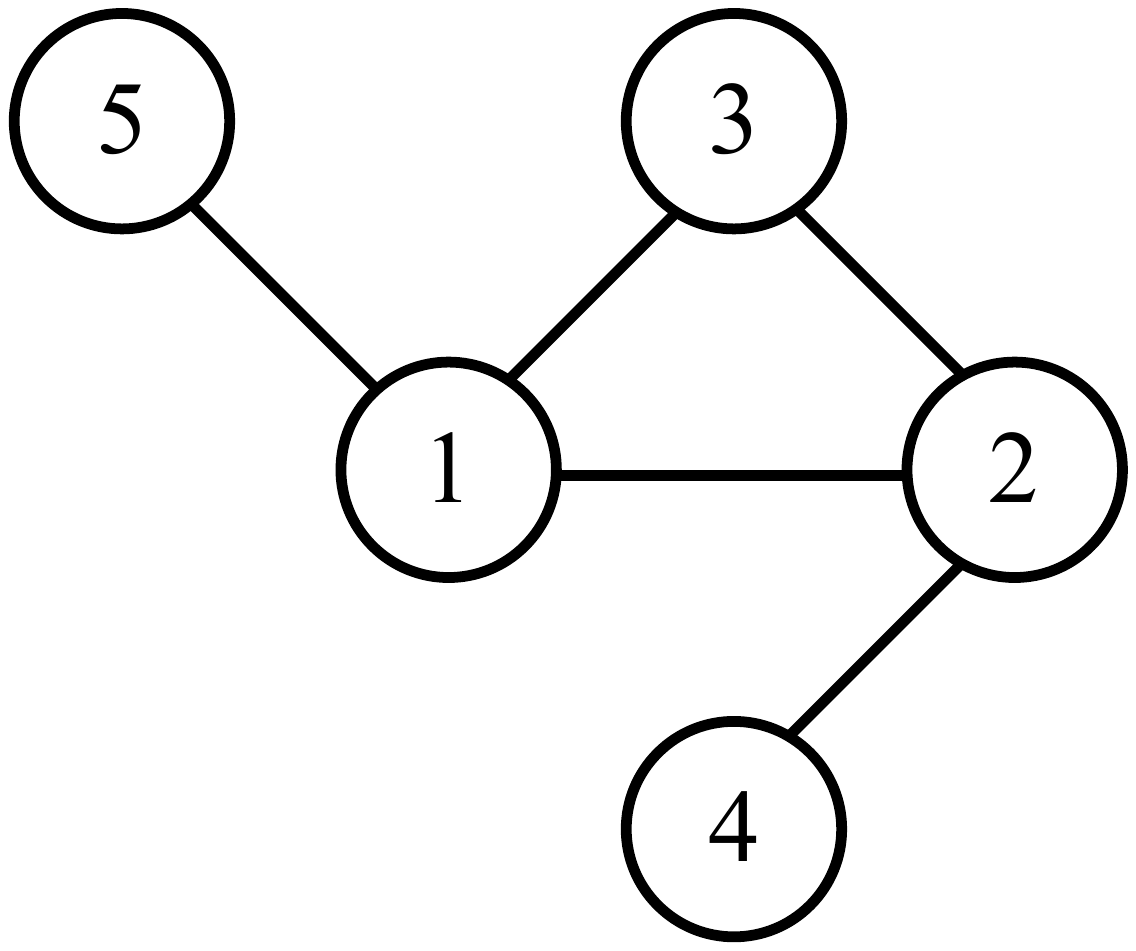}
\caption{\label{fig:subgraph}
A graph with $5$ vertices $V = \{1,\dots,5\}$ and $5$ edges $E=\{(12), (13),\dots, (24)\}$. 
The product $\prod_{(1k)\in E, k\neq 2}$ involves the vertices connected to $1$ that are not $2$, i.e.\ the vertices $3$ and $5$. 
The product $\prod_{(1k)\in E, (2k)\notin E}$ involves the vertices that are connected to $1$ but that are not connected to $2$, which is the single vertex $5$. (Note that it is implicit that $k=2$ should be excluded.)
The product $\prod_{(1k)\in E, (2k)\in E}$ involves the vertices that are connected to both $1$ and $2$, which is the vertex $3$. 
}
\end{figure}

Given an undirected graph $G = (V, E)$ defined by vertices $V=\{1,\dots,n\}$ and edges $E\subset V\times V$, as well an external field $h_i$ associated with each vertex, and a coupling strength (edge weight) $J_{ij}$ associated with each edge, we aim to find the configuration $s\in\{-1,+1\}^n$ that minimizes the (Ising) cost function
\begin{align}
C & =  \sum_{i\in V} h_i s_i + \sum_{(i,j)\in E} J_{ij} s_i s_j. 
\end{align}
This cost function is equivalent to the cost Hamiltonian
\begin{align}
H_C &= \sum_{i\in V} C_i + \sum_{(i,j)\in E} C_{ij} 
\end{align}
where 
\begin{align}
\left\{
\begin{aligned} 
C_{i} & = h_i \sigma^z_i \\
C_{ij} & = J_{ij} \sigma^z_i \sigma^z_j
\end{aligned}
\right.
.
\end{align}
Here $\sigma^x_i,\sigma^y_i,\sigma^z_i$ correspond to the Pauli matrices. The unitary transformation $U$ for a \QAOA circuit with depth $p=1$ is defined as
\begin{align}
U & = U_B^\beta U_C^\gamma  
\end{align}
where
\begin{align}
U_C^\gamma & = e^{-\imag\gamma H_C}= e^{-\imag\gamma (\sum_{ij} J_{ij} \sigma^z_i \sigma^z_j + \sum h_i \sigma^z_i)} \\
& = \prod_{(i,j)\in E} (\cos(J_{ij} \gamma)\sigma_i^0\sigma_j^0 - \imag \sin( J_{ij}\gamma) \sigma^z_i \sigma^z_j) \prod_{i\in V} (\cos(h_i \gamma)\sigma_i^0 - \imag \sin(h_i\gamma) \sigma^z_i) \\
U_B^\beta & = e^{-\imag\beta H_B}= e^{-\imag\beta \sum_i \sigma^x_i} \\
& = \prod_{i\in V} ( \cos(\beta)\sigma_i^0 - \imag \sin(\beta) \sigma^x_i )  .
\end{align}
Applying $U = U_B^\beta U_C^\gamma$ to an initial, uniform superposition of all bit strings, $\ket{\psi_0}$, we obtain the final \QAOA state
\begin{align}
\ket{\beta,\gamma}& =U \ket{\psi_0}.
\end{align}
The expectation value of $H_C$ for this final state is
\begin{align}
F(\beta, \gamma) & \coloneqq \sum_{i\in V} \langle C_{i} \rangle  + \sum_{(i,j)\in E} \langle C_{ij} \rangle 
\end{align}
where
\begin{align}
\left\{
\begin{aligned}
\langle C_{i} \rangle  
& = \bra{\beta,\gamma}C_i\ket{\beta,\gamma} 
= \bra{\psi_0}U^\dagger C_i U \ket{\psi_0} 
= h_i \Tr [\rho_0 U^{\dagger} \sigma^z_{i} U]\\
\langle C_{ij} \rangle 
& = \bra{\beta,\gamma}C_{ij}\ket{\beta,\gamma} 
= \bra{\psi_0}U^\dagger C_{ij} U \ket{\psi_0} 
= J_{ij}\Tr [\rho_0 U^{\dagger} \sigma^z_i\sigma^z_j U]\\
\end{aligned}
\right.
\label{eq:trace_ci}
\end{align}
and $\rho_0 =\ket{\psi_0} \bra{\psi_0} $ is the initial density matrix. The expectation value can be computed classically, however, a quantum computer is still required to obtain the bitstring with the relevant energy. 

\subsection{Formula for General Case with Variable Couplings}
In Appendix~\ref{app:traces} we derive explicit expressions for the two traces of Equation~\ref{eq:trace_ci}, giving us the results 
\begin{align} 
\left\{\begin{aligned}
\label{eq:main_res}
\langle C_i \rangle & = 
h_i \sin(2\beta)\sin(2\gamma h_i)\prod_{(ik)\in E}\cos(2\gamma J_{ik}) \\
\begin{split}
\langle C_{ij} \rangle & = 
\frac{J_{ij} \sin(4\beta)}{2} \sin(2 \gamma J_{ij}) 
\Big[ \cos(2 \gamma h_i) \prod_{\substack{(ik) \in E\\ k \neq j}} \cos(2 \gamma J_{ik} ) + \cos(2 \gamma h_j) \prod_{\substack{(jk) \in E \\ k \neq i}} \cos(2 \gamma J_{jk})\Big]  \\
& \quad -\frac{J_{ij}}{2} (\sin(2\beta))^2  \prod_{\substack{(ik)\in E\\ (jk)\notin E}} \cos(2 \gamma J_{ik})  \prod_{\substack{(jk)\in E\\ (ik)\notin E}}\cos(2\gamma J_{jk})  \\
& \qquad \times \Big[\cos(2 \gamma (h_i{+}h_j)) \prod_{\substack{(ik)\in E\\(jk)\in E}} \cos(2 \gamma (J_{ik}{+}J_{jk})) - 
\cos(2 \gamma (h_i{-}h_j)) \prod_{\substack{(ik)\in E\\(jk)\in E}} \cos(2 \gamma (J_{ik}{-}J_{jk}))\Big].
\end{split}
\end{aligned}
\right.
\end{align}
Equation~\ref{eq:main_res} is the contribution of a single vertex and connected subgraphs in the graph. Connected subgraphs are formed by two vertices $(i,j)$ connected by an edge and the nearest neighbours of the $i$ and $j$ vertices, as shown in Fig.~\ref{fig:subgraph}. From this general formula we will derive the expressions corresponding to simpler problem instances and graphs, including previously derived results.

Without loss of generality we can assume the graph to the complete graph $K_n$ as we can use $J_{ij}=0$ for absent edges. 
Doing so simplifies the above expressions  to 
\begin{align} 
\left\{\begin{aligned}
\langle C_i \rangle_K & = 
h_i \sin(2\beta)\sin(2\gamma h_i)\prod_{k\neq i}\cos(2\gamma J_{ik}) \\
\begin{split}
\label{eq:sk_h}
\langle C_{ij} \rangle_K & =  
\frac{J_{ij} \sin(4\beta)}{2} \sin(2 \gamma J_{ij}) 
\Big[ \cos(2 \gamma h_i) \prod_{k \neq i,j} \cos(2 \gamma J_{ik} ) + \cos(2 \gamma h_j) \prod_{k \neq i,j} \cos(2 \gamma J_{jk})\Big]  \\
& \quad -\frac{J_{ij}}{2} (\sin(2\beta))^2 \Big[\cos(2 \gamma (h_i{+}h_j)) \prod_{k \neq i,j} \cos(2 \gamma (J_{ik}{+}J_{jk})) - 
\cos(2 \gamma (h_i{-}h_j)) \prod_{k \neq i,j} \cos(2 \gamma (J_{ik}{-}J_{jk}))\Big].
\end{split}
\end{aligned}
\right.
\end{align}
If, additionally, one assumes the absence of local fields, i.e., $h_i=0$ and $h_j=0$, this further simplifies to 
\begin{align} 
\begin{split}
\label{eq:sk_h0}
\langle C_{ij} \rangle_{K,h=0} & =  
\frac{J_{ij} \sin(4\beta)}{2}\sin(2 \gamma J_{ij}) 
\Big[  \prod_{k \neq i,j} \cos(2 \gamma J_{ik} ) + \prod_{k \neq i,j} \cos(2 \gamma J_{jk})\Big]  \\
& \quad -\frac{J_{ij}}{2} (\sin(2\beta))^2 \Big[ \prod_{k \neq i,j} \cos(2 \gamma (J_{ik}{+}J_{jk})) - 
 \prod_{k \neq i,j} \cos(2 \gamma (J_{ik}{-}J_{jk}))\Big].
\end{split}
\end{align}
This last expression has been derived previously and appears in Lemma~C.1 of \cite{Bravyi2019ObstaclesTS}.

Returning to the general Equation~\ref{eq:main_res}, we can also consider the case of triangle-free graphs, i.e.,  graphs where $i$ and $j$ share no common vertices $k$ with both $(i,k)\in E$ and $(j,k)\in E$, for which the expectation reduces to
\begin{align}
\begin{split}
\label{eq:main_res2_notrian]}
\langle C_{ij} \rangle_{\mathrm{tf}} & = 
\frac{J_{ij} \sin(4\beta)}{2} \sin(2 \gamma J_{ij})  \Big[ \cos(2 \gamma h_i) \prod_{\substack{(ik) \in E\\ k \neq j}} \cos(2 \gamma J_{ik} ) + \cos(2 \gamma h_j) \prod_{\substack{(jk) \in E\\ k \neq i}} \cos(2 \gamma  J_{jk})\Big]  \\
& \quad +J_{ij} (\sin(2\beta))^2 \sin(2 \gamma h_i) \sin(2 \gamma h_j)
\prod_{\substack{(ik)\in E\\ (jk)\notin E}} \cos(2 \gamma J_{ik})  \prod_{\substack{(jk)\in E\\ (ik)\notin E}}\cos(2\gamma J_{jk}).
\end{split}
\end{align}

\subsection{Formulae for Ising Problems with Constant Couplings}
Next, we will derive the expressions of the expectation value for some special-case Hamiltonians where the couplings are constant ($J_{ij}{=}\pm 1$). For the NP-hard ``P5'' Ising problem of \cite{Barahona_1982} for a graph $(V,E)$ the goal is to minimize the Hamiltonian 
\begin{align}
    H_{\mathrm{P5}} &= \sum_{(i,j)\in E} \sigma^z_i \sigma^z_j + \sum_{i\in V} \sigma^z_i 
\end{align}
With $d_i$ and $d_j$ the degrees of $i$ and $j$, and $f_{ij}$ the number of shared vertices between $i$ and $j$, it is easy to see that for the expectation we now have from Equation~\ref{eq:main_res}
\begin{align}
\label{eq:Cij_P5}
\left\{
\begin{aligned}
\langle C_{i} \rangle_{\mathrm{P5}} 
& = \sin(2\beta) \sin(2\gamma)(\cos(2\gamma))^{d_i} \\
\begin{split}
\langle C_{ij} \rangle_{\mathrm{P5}}
& =  
\frac{\sin(4\beta)}{2} \sin(2 \gamma ) 
\Big[ (\cos(2 \gamma ))^{d_i} + (\cos(2 \gamma))^{d_j}\Big]\\
& \quad + \frac{1}{2} (\sin(2\beta))^2  (\cos(2 \gamma ))^{d_i+d_j-2f_{ij}-2}  (1-(\cos(4 \gamma ))^{f_{ij}+1}).
\end{split}
\end{aligned}\right.
\end{align}

For the \MaxCut problem, with no local fields and couplings $J_{ij}=-1$, the expectation reads
\begin{align}
\left\{
\begin{aligned}
H_{\mathrm{MC}}& =-\sum_{(i,j)\in E} \sigma^z_i \sigma^z_j  \\ 
\langle C_{i} \rangle_{\mathrm{MC}} & = 0 \\ 
\langle C_{ij} \rangle_{\mathrm{MC}} & 
= \frac{\sin(4\beta)}{2} \sin(2\gamma) [(\cos(2\gamma))^{d_i-1}+(\cos(2\gamma))^{d_j-1}] \\
& \quad -\frac{1}{2}(\sin(2\beta))^2 (\cos(2 \gamma))^{d_i+d_j-2f_{ij}-2} (1-(\cos(4\gamma))^{f_{ij}}) .
\end{aligned}
\right.
\end{align}
This is the result obtained by Wang et al.\ \cite{wang2018quantum}. 

The expression in the absence of spin-spin couplings reads,
\begin{align}
\label{eq:main_result_nocoupling1}
H_{\mathrm{NC}} & = \sum_{i\in V}h_i  \sigma^z_i  \\
\langle C_{i} \rangle_{\mathrm{NC}} & = 
h_i \sin(2\beta)  \sin(2\gamma h_i) \label{eq:no_coupling} \\
\label{eq:main_result_nocoupling2}
\langle C_{ij} \rangle_{\mathrm{NC}} & = 0.    
\end{align}
The solution to this optimization problem is trivial, however, the determination of the angles for the optimal expectation value is not.

We will use the analytical formulae derived in this section to gain insight into the behavior of expectation value landscapes for \QAOA $p=1$. First, we will study \MaxCut instances, as the paper progresses we will gradually increase the complexity of the instances we study until we reach general Ising problems.

\subsection{Formulae for Standard Sherrington-Kirkpatrick Model without Local Field}
\label{sec:SK_maxcut_expectation}
To analyze the Sherrington-Kirkpatrick model, Equation~\ref{eq:sk_h0},  we assume  no local field, i.e.\ $h_i=0$, and that the all-to-all coupling values $J_{ij}$ are distributed according to an independent and identical distribution with mean $0$. 
Taking the average over this distribution and using the facts that all $J_{ij}$ distributions are independent and that $\EE_J[J_{ij}]=0$ we are interested in the expectation of expectations 
\begin{align} 
\EE_J[\langle C_{ij} \rangle] & =  
\sin(4\beta)\cdot \EE_J[J  \sin(2 \gamma J)] 
\cdot(\EE_J[\cos(2 \gamma J)])^{n-2}.
\end{align}
Often one is interested in the $n\rightarrow \infty$ behaviour of this model, which is easiest to discuss when we scale the $J_{ij}$ random variables by $1/\sqrt{n-1}$ such that we have 
\begin{align}
J_{ij} & = \frac{J'}{\sqrt{n-1}},
\end{align}
where $J'$ is a random variable with mean $0$ and 
variance $\sigma^2 \in O(1)$ independent of $n$. 
Doing so will result in an $O(1)$ energy per spin,  i.e.\ $\langle C/n\rangle = (n-1)/2\cdot \langle C_{ij}\rangle$ will be of order $1$.  
We use $\sqrt{n-1}$ instead of the more common $\sqrt{n}$ to make the finite $n$ expressions work out nicer. 

For the normal distribution $J'\sim N(0,\sigma^2)$ we have $J\sim N(0,\sigma^2/(n-1))$, and hence   
\begin{align}
\EE_{J\sim N(0,\sigma^2/(n-1))}[J\sin(2\gamma J)] & = 2\gamma\sigma^2\cdot (n-1)^{-1}\cdot e^{-2\gamma^2\sigma^2/(n-1)}\\ 
\EE_{J \sim N(0,\sigma^2/(n-1))}[\cos(2\gamma J)] & = e^{-2\gamma^2\sigma^2/(n-1)}. 
\end{align}
The expectation of the expectations of the energy per spin therefore equals
\begin{align} \label{eq:expSK}
\EE_{J'\sim N(0,\sigma^2)}[\langle C/n \rangle] & =  
\sin(4\beta)\cdot\gamma \sigma^2 e^{-2\gamma^2\sigma^2}.
\end{align}
Note that this is an exact expression for finite $n$ and it allows us to compute the optimal angles $\beta$, $\gamma$ that minimize the expectation as a function of $n$ and $\sigma$:  
\begin{align} \label{eq:expSKopt}
\beta_\textrm{min} = \frac{-\pi}{8}, \quad 
\gamma_\textrm{min} =  \frac{1}{2\sigma}, \quad 
\text{with}\quad
\EE_{J'\sim N(0,\sigma^2)}[\langle C/n \rangle]  = - \frac{\sigma}{2\sqrt{e}} \approx - 0.303 \sigma.
\end{align}

All of the above results agree with those in \cite[Section~4]{farhi2019quantum}. Note, however, that the authors in \cite{farhi2019quantum} take the equivalent approach of normalizing the entire cost function by the $\sqrt{n}$ factor, not just $J$.
Moreover, \cite{farhi2019quantum} shows that in the $n\rightarrow \infty$ limit this result is independent of the details of the distribution $J'$, given that it has mean 0 and variance 1.   
As an example, consider the bimodal distribution $B$ where $\Pr[J'{=}{\pm}\sigma] = 1/2$ for some fixed $\sigma$, such that $\EE[J']=0$ and $\EE[J'^2] = \sigma^2$. 
As a result,  we have $\Pr[J{=}{\pm}\sigma/\sqrt{n-1}] = 1/2$ with $\EE_{J}[J \sin(2\gamma J)] = (\sigma/\sqrt{n-1})\sin(2\gamma \sigma/\sqrt{n-1})$ and $\EE_{J}[\cos(2\gamma J)] = \cos(2\gamma \sigma/\sqrt{n-1})$, and hence
\begin{align}
\EE_{J'\sim B}[\langle C/n\rangle] & = 
\frac{1}{2}\sin(4\beta) \cdot 
\sigma\sqrt{n-1}\cdot \sin(2\gamma \sigma/\sqrt{n-1})\cdot(\cos(2\gamma \sigma/\sqrt{n-1}))^{n-2}. 
\end{align}
In the $n\rightarrow \infty$ limit for large systems we thus get again 
\begin{align}
\lim_{n\rightarrow \infty}\EE_{J'\sim B}[\langle C/n \rangle] & = 
\sin(4\beta)\cdot \gamma \sigma^2 e^{-2\gamma^2 \sigma^2}, 
\end{align}
which is identical to Equation~\ref{eq:expSK}. 

It is important to realize, though, that there are distributions over $J'$ with mean $0$ and variance $\sigma^2$ that give different results for the optimal angles and expectations.  
Consider the setting where we want to describe a spin glass where each vertex has an expected number of $d$ edges for which it has $J{=}{\pm}1$, while for the other $n-1-d$ edges it has $J{=}0$. 
To capture this setting  we define the random variable $J$ according to a trimodal distribution $T$ over coupling values $J{\in} \{-1,0,+1\}$ with $\Pr[J{=}{\pm}1] = d/(2n-2)$ and $\Pr[J{=}0]=1-d/(n-1)$ such that $\EE_{J\sim T}[J]=0$ and $\EE_{J\sim T}[J^2]= d/(n-1)$.
For this distribution we have the expectations $\EE_{J\sim T}[J\sin(2\gamma J)] = d/(n-1) \sin(2\gamma)$, and $\EE_{J\sim T}[\cos(2\gamma J)] = 1+d(\cos(2\gamma)-1)/(n-1)$ and hence
\begin{align}
\EE_{J\sim T}[\langle C/n\rangle] & = 
\frac{1}{2}\sin(4\beta) \cdot  d \cdot \sin(2\gamma) \cdot (1-d/(n-1)+d/(n-1)\cos(2\gamma))^{n-2}. 
\end{align}
In the $n\rightarrow \infty$ limit this gives 
\begin{align}\label{eq:SK-d-n-inf}
\EE_{J\sim T}[\langle C/n \rangle] & = 
\frac{1}{2}\sin(4\beta)\cdot d \cdot \sin(2\gamma)\cdot e^{-2d(\sin(\gamma))^2}.
\end{align}

If we set $d=1$ the Equation~\ref{eq:SK-d-n-inf} is maximized by $\beta = -\pi/8$, $\gamma = 0.452278$, such that $\EE_{J\sim T}[\langle C/n\rangle] = -0.268281$. 
Note now that this derived optimum is different from the results of Equation~\ref{eq:expSKopt}. 
This difference is explained by the fact that the related distribution of $J' = J\sqrt{n-1}$ (with mean $0$ and variance $1$) is no longer independent of $n$, as it is defined by $\Pr[J'{=}{\pm}\sqrt{n-1}] = 1/(2n-2)$ and $\Pr[J'{=}0] = (n-2)/(n-1)$. 
Because of this $n$-dependency it no longer holds that the $n\rightarrow \infty$ properties are described by  Equations~\ref{eq:expSK} and \ref{eq:expSKopt}.

More specifically,  Equation~\ref{eq:expSK} will hold as long as we have the expectations in the $n\rightarrow \infty$ limit:
\begin{align}\label{eq:Jmoments}
 \EE_J[nJ^2] \rightarrow \sigma^2\text{, and~}
 \EE_J[nJ^r] \rightarrow 0, \text{~for $r \in \{3,4,\dots\}$.} 
\end{align}
That way we have 
\begin{align} \label{eq:ninfinitylimit}
\lim_{n\rightarrow \infty}\EE_J[\langle C/n\rangle] & = 
\lim_{n\rightarrow \infty}\frac{n-1}{2}\cdot\sin(4\beta)\cdot \EE_J[J\sin(2\gamma J)]\cdot(\EE_J[\cos(2\gamma J)])^{n-2} \\
& = \sin(4\beta)\cdot \gamma\sigma^2 \cdot e^{-2\gamma^2 \sigma^2}.
\end{align}
The above trimodal distribution does not fit this mold as it has $\EE_J[nJ^3] \rightarrow d$.

\subsection{Formulae for the Sherrington-Kirkpatrick Model with Constant Local Field}
Consider now a variation of the SK model with a constant local field $h$ and with $J_{ij}$ couplings that again are i.i.d.\ with mean zero, the case shown in Equation~\ref{eq:sk_h}. The local energies can now be expressed as 
\begin{align} 
\left\{\begin{aligned}
\EE_J[\langle C_i \rangle] & = 
h \sin(2\beta)\sin(2\gamma h)\cdot \EE_J[\cos(2\gamma J)]^{n-1} \\
\EE_J[\langle C_{ij} \rangle] & =  
\sin(4\beta)\cos(2 \gamma h)\cdot \EE_J[J  \sin(2 \gamma J)] \cdot \EE_J[\cos(2 \gamma J)]^{n-2}.
\end{aligned}
\right.
\end{align}
The per spin energy thus equals
\begin{align} 
\EE_J[\langle C/n\rangle] & = 
h \sin(2\beta)\sin(2\gamma h)\cdot \EE_J[\cos(2\gamma J)]^{n-1} + \frac{n-1}{2}\sin(4\beta)\cos(2 \gamma h)\cdot \EE_J[J  \sin(2 \gamma J)] \cdot \EE_J[\cos(2 \gamma J)]^{n-2}.
\end{align}

Assuming again that the $J$ values are normally distributed according to $J=J'/\sqrt{n-1}$ with $J'\sim N(0,\sigma^2)$ we get for the energy per spin
\begin{align}
\EE_{J'\sim N(0,\sigma^2)}[\langle C/n \rangle]
& = (\gamma \sigma^2 \cos(2 h \gamma) \sin(4 \beta) + h \sin(2 \beta) \sin(2 h \gamma))e^{-2 \gamma^2\sigma^2}.
\end{align}

As an example, if we take $\sigma=1$ and $h=1$ we have the optimal angles $\beta \approx -0.54854$ and $\gamma \approx 0.3962$, such that $\EE[\langle C/n\rangle] \approx -0.62791$. 
Compare this with previous result for $(\sigma,h) = (1,0)$ for which the optimal angles are $\beta = \frac{-\pi}{8} \approx -0.3927$ and $\gamma = 0.5$, with 
$\EE[\langle C/n\rangle]= - \frac{1}{2\sqrt{e}} \approx -0.3033$. This shows that the presence of a local field does change not only the optimal angles but the expectation values in a significant way.  

\subsection{Formulae for the Sherrington-Kirkpatrick Model with Normal Local Field}
Next, we will study the SK model with local fields distributed according to an independent and identical distribution with mean 0, again the case of Equation~\ref{eq:sk_h}. The local energies can be expressed as
\begin{align} 
\left\{\begin{aligned}
\EE_{J,h}[\langle C_i \rangle] & = 
 \sin(2\beta)\cdot \EE_h[h \sin(2\gamma h)]\cdot \EE_J[\cos(2\gamma J)]^{n-1} \\
\EE_{J,h}[\langle C_{ij} \rangle] & =  
\sin(4\beta)\cdot \EE_h[\cos(2 \gamma h)]\cdot \EE_J[J  \sin(2 \gamma J)] \cdot \EE_J[\cos(2 \gamma J)]^{n-2}.
\end{aligned}
\right.
\end{align}
The per spin energy thus equals
\begin{align} 
\EE_{J,h}[\langle C/n\rangle] & = 
\sin(2\beta)\cdot \EE_h[h \sin(2\gamma h)]\cdot \EE_J[\cos(2\gamma J)]^{n-1}\\& \quad + \frac{n-1}{2}\sin(4\beta)\cdot\EE_h[\cos(2 \gamma h)]\cdot \EE_J[J  \sin(2 \gamma J)] \cdot \EE_J[\cos(2 \gamma J)]^{n-2}.
\end{align}
Assume now that the  $h_i$ random variables will have the normal distribution $h\sim N(0,\sigma_h^2)$:
\begin{align}
 \EE_{h\sim N(0,\sigma_h^2)}[h\sin(2\gamma h)]  &= 2\gamma\sigma_h^2\cdot e^{-2\gamma^2\sigma_h^2}\\
\EE_{h \sim N(0,\sigma_h^2)}[\cos(2\gamma h)]  &= e^{-2\gamma^2\sigma_h^2}.
\end{align}
The expectation of the expectations of the energy per spin equals
\begin{align}
\EE_{\substack{J'\sim N(0,\sigma^2)\\ h \sim N(0,\sigma_h^2)}}[\langle C/n \rangle] & =  
(2 \sigma_h^2 \sin(2\beta) + \sigma^2 \sin(4\beta))\cdot \gamma e^{-2\gamma^2 (\sigma^2+\sigma_h^2)}.
\end{align}
We then compute the optimal angles $\beta$, $\gamma$,
\begin{align} 
\beta_\textrm{min} = -\arccos{(\frac{\sqrt{4-\frac{\sigma_h^2 + \sqrt{8 \sigma^4 +\sigma_h^4 }}{\sigma^2}}}{2\sqrt{2}})}, \quad 
\gamma_\textrm{min} =  \frac{1}{2 \sqrt{\sigma^2+\sigma_h^2}}
\end{align}

We can reduce the expression of the expectation of the expectations of the energy per spin by making both parameters have the same distribution $\sigma=\sigma_h$
\begin{align}
\EE_{\substack{J'\sim N(0,\sigma^2)\\ h \sim N(0,\sigma^2)}}[\langle C/n \rangle] & =  
(2 \sin(2\beta) + \sin(4\beta))\cdot \gamma \sigma^2 e^{-4\gamma^2\sigma^2}.
\end{align}
And obtain the optimal angles $\beta$, $\gamma$,
\begin{align} 
\beta_\textrm{min} = \frac{-\pi}{6}, \quad 
\gamma_\textrm{min} =  \frac{1}{2 \sqrt{2} \sigma}, \quad 
\text{with}\quad
\EE_{\substack{J'\sim N(0,\sigma^2)\\h \sim N(0,\sigma^2)}}[\langle C/n \rangle]  = - \frac{3 }{4} \sqrt{\frac{3}{2 e}} \sigma \approx -0.557 \sigma.
\end{align}

\subsection{Formulae for Random Spin Glasses on Regular Graphs}
\label{sec:reg_graphs_expectation}
When considering random $J$ couplings between vertices of a constant degree $d$ graph $G=(V,E)$ and no local field, the expectation of expectations for edges $(ij)\in E$ derived from Equation~\ref{eq:main_res} will be 
\begin{align} 
\EE_J[\langle C_{ij} \rangle] & =  
\sin(4\beta)\cdot \EE_J[J  \sin(2 \gamma J)] 
\cdot(\EE_J[\cos(2 \gamma J)])^{d-1},
\end{align}
where $d$ is the degree of the graph.
For non-edges $(ij)\notin E$ we obviously have $\EE_J[\langle C_{ij} \rangle] =  0$.
If we assume a normal distribution of $J$ couplings over the edges and for each vertex $i$ the expectation $\EE_J[\sum_j J_{ij}^2] = \sigma^2$, we should take $J\sim N(0,\sigma^2/d)$. 
With this distribution we get 
\begin{align}
\EE_{J\sim N(0,\sigma^2/d)}[J\sin(2\gamma J)] & = 2\gamma\sigma^2\cdot d^{-1}\cdot e^{-2\gamma^2\sigma^2/d}\\ 
\EE_{J \sim N(0,\sigma^2/d)}[\cos(2\gamma J)] & = e^{-2\gamma^2\sigma^2/d} 
\end{align}
and hence an expectation of the expectations of the energy per spin of
\begin{align}
\EE_{J\sim N(0,\sigma^2/d)}[\langle C/n \rangle] & =  
\sin(4\beta)\cdot \frac{n-1}{d}\cdot \gamma\sigma^2\cdot e^{-2\gamma^2\sigma^2}.
\end{align}
The corresponding optimal values are thus
\begin{align}
\beta_\textrm{min} = \frac{-\pi}{8}, \quad 
\gamma_\textrm{min} =  \frac{1}{2\sigma}, \quad 
\text{and}\quad
\EE_{J'\sim N(0,\sigma^2)}[\langle C/n \rangle]  = - \frac{n-1}{d}\cdot \frac{\sigma}{2\sqrt{e}}, 
\end{align}
which for $d=n-1$ returns the standard SK result. 

Next, we add local fields. We obtain for degree $d$ graphs with normal distribution $h\sim N(0,\sigma^2)$,
\begin{align} 
\left\{\begin{aligned}
\EE_{J,h}[\langle C_i \rangle] & = 
 \sin(2\beta)\cdot \EE_h[h \sin(2\gamma h)]\cdot \EE_J[\cos(2\gamma J)]^{d} \\
\EE_{J,h}[\langle C_{ij} \rangle] & =  
\sin(4\beta)\cdot \EE_h[\cos(2 \gamma h)]\cdot \EE_J[J  \sin(2 \gamma J)] \cdot \EE_J[\cos(2 \gamma J)]^{d-1}.
\end{aligned}
\right.
\end{align}

The expectation of the expectations of the energy per spin equals
\begin{align}
\EE_{\substack{J'\sim N(0,\sigma^2)\\h \sim N(0,\sigma^2)}}[\langle C/n \rangle] & =  
(2\sin(2\beta) + \frac{n-1}{d} \sin(4\beta))\cdot \gamma \sigma^2 e^{-4\gamma^2\sigma^2}.
\end{align}

Here the optimal angle $\gamma_\textrm{min}$ remains constant as we modify $d$, while $\beta_\textrm{min}$ changes.

\section{Numerical Results and discussion}

\begin{table}[t]
\begin{tabular}{c c c c c}
{Figure} & {$h_i$} & {$J_{ij}$} & {n} & d \\ \hline \hline
Fig. 2 & 0 & $\{-1, 1\}$ & 23~/~14~/~11 & Sycamore~/~3~/~10 \\ \hline
Fig. 3 & 0 & $\{-1, 1\}$ & 100 & 3~/~99 \\ \hline
Fig. 4 & 0 & $\{-1, 1\}$ & 7000 & 11.85 \\ \hline
Fig. 5 & $\{-1, 1\}$ & $\{-1, 1\}$ & 23~/~14~/~11 & Sycamore~/~3~/~10 \\ \hline
Fig. 6 & $\{-2, 2\}$ & $\{-2, 2\}$ & 23~/~14~/~11 & Sycamore~/~3~/~10 \\ \hline
Fig. 7 & $\{-2, 2\}$ & $\{-1, 1\}$ & 23~/~14~/~11 & Sycamore~/~3~/~10 \\ \hline
Fig. 8 & $\{-10, 10\}$ & $\{-1, 1\}$ & 23~/~14~/~11 & Sycamore~/~3~/~10 \\ \hline
Fig. 9 & $\{-1, 1\}$ & $\{-2, 2\}$ & 23~/~14~/~11 & Sycamore~/~3~/~10 \\ \hline
Fig. 10 & $\{-1, 1\}$ & $\{-10, 10\}$ & 23~/~14~/~11 & Sycamore~/~3~/~10 \\ \hline
Fig. 11 & $\{-30, -29,\dots,29,30\}$ & $\{-30,30\}$ & 16 & 3 \\ \hline
\end{tabular}
\caption{Summary of the parameters of the figures present in the text.}
\label{table:figure_summary}
\end{table}

\begin{figure}[t]
\centering
\includegraphics[width=0.6\textwidth]{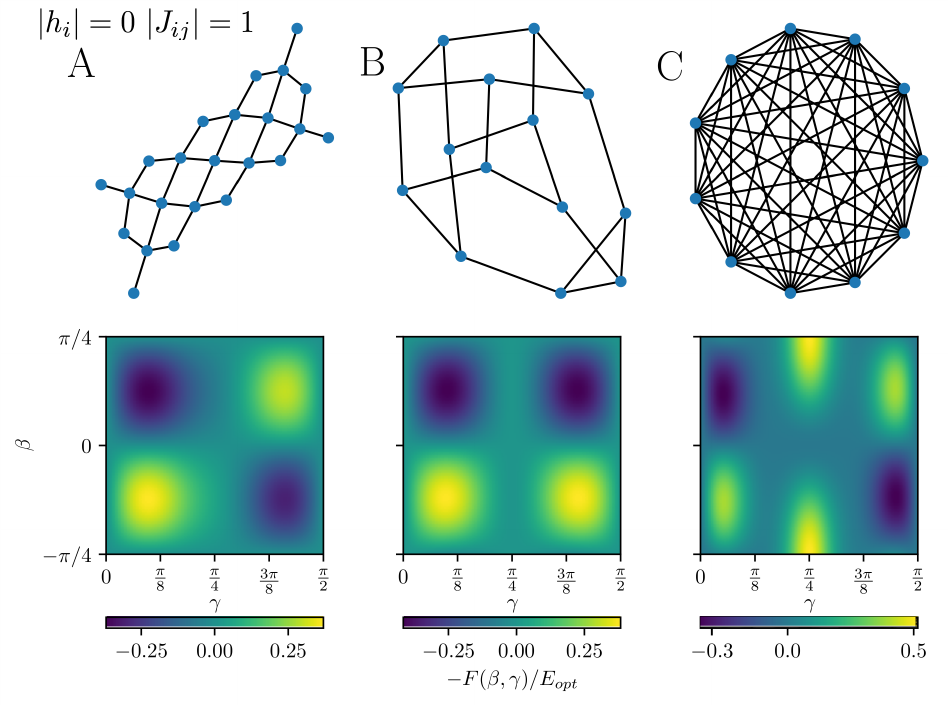}
\caption{\label{fig:google_plots} Negative expectation-value landscapes, obtained using our analytical formulae, Eq.~\ref{eq:main_res}, for the same problem instances studied experimentally in \cite{google_qaoa}. Normalized by the optimal energy $E_\textrm{opt}$. The graphs for each instance are shown above their respective landscape. 
(A) Sycamore grid problem with $23$ qubits. 
(B) $3$-regular graph with $14$ qubits. 
(C) Fully connected SK-model with $11$ qubits.}
\end{figure}

\begin{figure}[t]
\centering
\includegraphics[width=0.6\textwidth]{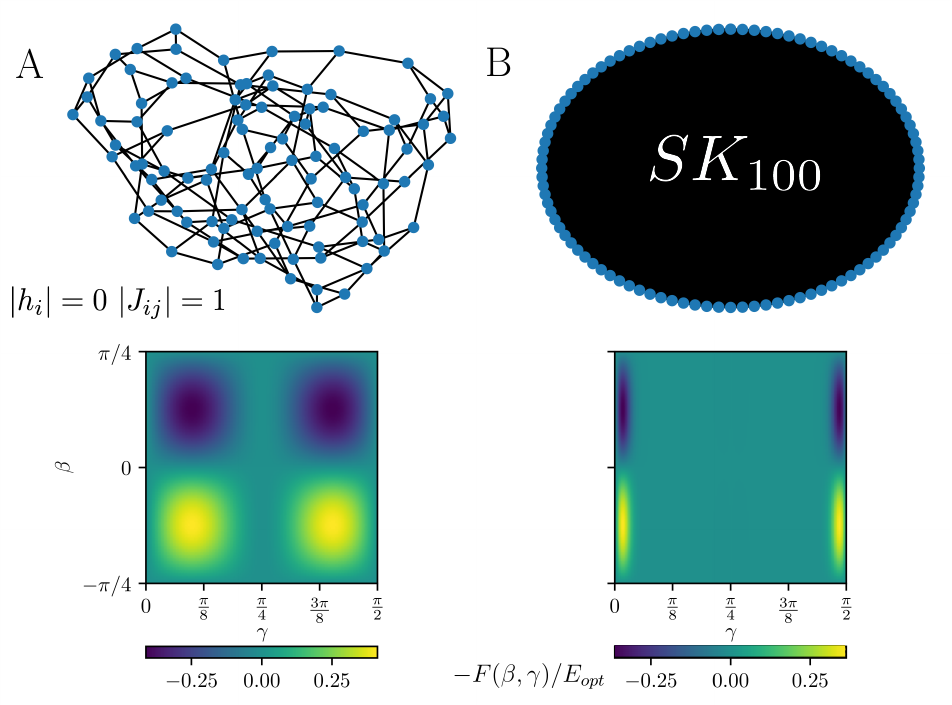}
\caption{\label{fig:google_plots_morequbits} Negative expectation-value landscapes for two $100$-qubit instances. Normalized by the optimal energy $E_\textrm{opt}$. 
(A) 3-regular graph with $100$ qubits. 
(B) Fully connected SK-model with $100$ qubits.}
\end{figure}

\begin{table}[t]
 \centering
 \begin{tabular}{c | S[table-format=4.0] S[table-format=5.0] S[table-format=5.0] S[table-format=5.0] S[table-format=4.3] S[table-format=1.3] S[table-format=5.3] S[table-format=1.3] c}
 Name & 	{$|V|$}    & {$|E|$}     & {Sum of weights} & {Best-known Ising} & {\QAOA expectation} & {Ratio Ising} & {$1.18\sqrt{|E| \times |V|}$}    & {Ratio exp.} & Weights \\ \hline\hline
 G1   & 800  & 19176 & 19176          & 4072             & 1482.034         & 0.364       & 4621.745  & 0.321 & $\{0,1\}$     \\ \hline
 G2   & 800  & 19176 & 19176          & 4064             & 1480.775          & 0.364       & 4621.745  & 0.32 & $\{0,1\}$     \\ \hline
 G3   & 800  & 19176 & 19176          & 4068             & 1478.236         & 0.363       & 4621.745  & 0.32  & $\{0,1\}$    \\ \hline
 G4   & 800  & 19176 & 19176          & 4116             & 1481.303         & 0.36       & 4621.745  & 0.321 & $\{0,1\}$      \\ \hline
 G5   & 800  & 19176 & 19176          & 4086             & 1479.496         & 0.362       & 4621.745  & 0.32 & $\{0,1\}$     \\ \hline
 G6   & 800  & 19176 & 154            & 4202             & 1679.595         & 0.4       & 4621.745  & 0.363 & $\{0,\pm 1\}$     \\ \hline
 G7   & 800  & 19176 & -150           & 4162             & 1674.41         & 0.402       & 4621.745  & 0.362 & $\{0,\pm 1\}$      \\ \hline
 G8   & 800  & 19176 & -170           & 4180             & 1676.846         & 0.401       & 4621.745  & 0.363 & $\{0,\pm 1\}$      \\ \hline
 G9   & 800  & 19176 & -64            & 4172             & 1676.17        & 0.402       & 4621.745  & 0.363  & $\{0,\pm 1\}$     \\ \hline
 G10  & 800  & 19176 & -160           & 4160             & 1675.396        & 0.403       & 4621.745  & 0.363  & $\{0,\pm 1\}$     \\ \hline
 G11  & 800  & 1600  & 34             & 1094             & 519.61           & 0.475       & 1335.018  & 0.389 & $\{0,\pm 1\}$      \\ \hline
 G12  & 800  & 1600  & -4             & 1116             & 519.61          & 0.466       & 1335.018  & 0.389 & $\{0,\pm 1\}$      \\ \hline
 G13  & 800  & 1600  & 34             & 1130             & 519.61           & 0.46        & 1335.018  & 0.389 & $\{0,\pm 1\}$      \\ \hline
 G14  & 800  & 4694  & 4694           & 1434             & 577.546          & 0.403      & 2286.644  & 0.253 & $\{0,1\}$     \\ \hline
 G15  & 800  & 4661  & 4661           & 1439             & 575.252         & 0.4        & 2278.592  & 0.252 & $\{0,1\}$     \\ \hline
 G16  & 800  & 4672  & 4672           & 1432             & 574.294          & 0.401       & 2281.279  & 0.252 & $\{0,1\}$     \\ \hline
 G17  & 800  & 4667  & 4667           & 1427             & 573.562          & 0.402      & 2280.058  & 0.252 & $\{0,1\}$     \\ \hline
 G18  & 800  & 4694  & 64             & 1920             & 714.097          & 0.372      & 2286.644  & 0.312 & $\{0,\pm 1\}$     \\ \hline
 G19  & 800  & 4661  & -113           & 1925             & 710.511          & 0.369       & 2278.592  & 0.312 & $\{0,\pm 1\}$     \\ \hline
 G20  & 800  & 4672  & -46            & 1928             & 709.332          & 0.368       & 2281.279  & 0.311 & $\{0,\pm 1\}$     \\ \hline
 G21  & 800  & 4667  & -67            & 1929             & 710.908          & 0.369       & 2280.058  & 0.312 & $\{0,\pm 1\}$     \\ \hline
 G27  & 2000 & 19990 & -42            & 6724             & 2693.905         & 0.401         & 7461.109  & 0.361 & $\{0,\pm 1\}$     \\ \hline
 G28  & 2000 & 19990 & -104           & 6700             & 2695.314         & 0.402       & 7461.109  & 0.361 & $\{0,\pm 1\}$     \\ \hline
 G29  & 2000 & 19990 & 80             & 6730             & 2693.685         & 0.4         & 7461.109  & 0.361 & $\{0,\pm 1\}$     \\ \hline
 G30  & 2000 & 19990 & 100            & 6726             & 2693.264          & 0.4       & 7461.109  & 0.361 & $\{0,\pm 1\}$     \\ \hline
 G31  & 2000 & 19990 & -80            & 6700             & 2691.868         & 0.402       & 7461.109  & 0.361 & $\{0,\pm 1\}$     \\ \hline
 G59  & 5000 & 29570 & 96             & 12076            & 4452.877          & 0.369       & 14348.043 & 0.31 & $\{0,\pm 1\}$      \\ \hline
 G61  & 7000 & 17148 & 362            & 11230            & 4590.429         & 0.409       & 12928.191 & 0.355 & $\{0,\pm 1\}$     \\ \hline
 G64  & 7000 & 41459 & 527            & 16975            & 6235.328         & 0.367       & 20102.054 & 0.31 & $\{0,\pm 1\}$      \\ \hline
 \end{tabular}
\caption{Computational results of \QAOA $p=1$ expectation values in Ising energies for a subset of the G-set instances. 
Column \emph{Sum of weights} shows the sum of all $J_{ij}$ values. 
Column \emph{Best-known Ising} shows the best classical results for the Ising energy, column \emph{\QAOA expectation} gives the best \QAOA expectation value. 
The best-known classical results were obtained from \cite{toshiba_benchmarking}. 
The columns \emph{Ratio Ising} and \emph{Ratio exp.} show the corresponding ratio of the \QAOA value versus the best classical result and versus the analytical formula value, respectively.}
\label{table:G-set_Ising}
\end{table}

\begin{table}[t]
\begin{tabular}{c | S[table-format=4.0] S[table-format=5.0] S[table-format=5.0] S[table-format=5.0] S[table-format=5.3] S[table-format=0.3] S[table-format=5.0] c}
Name & {$|V|$} & {$|E|$}  & {Sum of weights} & {Best-known cut} & {\QAOA expectation cut} & {Ratio cut} &  {Time (s)} & {Weights} \\ \hline\hline
G1   & 800  & 19176 & 19176          & 11624          & 10329.017                  & 0.889 & 28322 & $\{0,1\}$     \\ \hline
G2   & 800  & 19176 & 19176          & 11620          & 10328.387                   & 0.889 & 28562 & $\{0,1\}$     \\ \hline
G3   & 800  & 19176 & 19176          & 11622          & 10327.118                  & 0.889 & 28332 & $\{0,1\}$     \\ \hline
G4   & 800  & 19176 & 19176          & 11646          & 10328.651                  & 0.887 & 28430 & $\{0,1\}$     \\ \hline
G5   & 800  & 19176 & 19176          & 11631          & 10327.748                  & 0.888 & 29039 & $\{0,1\}$     \\ \hline
G6   & 800  & 19176 & 154            & 2178           & 916.797                    & 0.42 & 31953 & $\{0,\pm 1\}$      \\ \hline
G7   & 800  & 19176 & -150           & 2006           & 762.205                     & 0.38 & 32223 & $\{0,\pm 1\}$     \\ \hline
G8   & 800  & 19176 & -170           & 2005           & 753.423                    & 0.376 & 32685 & $\{0,\pm 1\}$     \\ \hline
G9   & 800  & 19176 & -64            & 2054           & 806.085                    & 0.392 & 33056 & $\{0,\pm 1\}$    \\ \hline
G10  & 800  & 19176 & -160           & 2000           & 757.698                    & 0.379 & 33681 & $\{0,\pm 1\}$    \\ \hline
G11  & 800  & 1600  & 34             & 564            & 276.805                    & 0.491 & 1193 & $\{0,\pm 1\}$    \\ \hline
G12  & 800  & 1600  & -4             & 556            & 257.805                    & 0.464 & 1188 & $\{0,\pm 1\}$     \\ \hline
G13  & 800  & 1600  & 34             & 582            & 276.805                    & 0.476 & 1194 & $\{0,\pm 1\}$     \\ \hline
G14  & 800  & 4694  & 4694           & 3064           & 2635.773                   & 0.86 & 4808 & $\{0,1\}$     \\ \hline
G15  & 800  & 4661  & 4661           & 3050           & 2618.126                   & 0.858 & 4982 & $\{0,1\}$     \\ \hline
G16  & 800  & 4672  & 4672           & 3052           & 2623.147                   & 0.859 & 5076 & $\{0,1\}$     \\ \hline
G17  & 800  & 4667  & 4667           & 3047           & 2620.281                   & 0.86 & 5056 & $\{0,1\}$      \\ \hline
G18  & 800  & 4694  & 64             & 992            & 389.049                    & 0.392 & 5125 & $\{0,\pm 1\}$     \\ \hline
G19  & 800  & 4661  & -113           & 906            & 298.755                    & 0.33 & 5187 & $\{0,\pm 1\}$     \\ \hline
G20  & 800  & 4672  & -46            & 941            & 331.666                    & 0.352 & 5186 & $\{0,\pm 1\}$     \\ \hline
G21  & 800  & 4667  & -67            & 931            & 321.954                    & 0.346 & 5237 & $\{0,\pm 1\}$     \\ \hline
G27  & 2000 & 19990 & -42            & 3341           & 1325.952                   & 0.397 & 22103 & $\{0,\pm 1\}$     \\ \hline
G28  & 2000 & 19990 & -104           & 3298           & 1295.657                   & 0.393 & 21916 & $\{0,\pm 1\}$     \\ \hline
G29  & 2000 & 19990 & 80             & 3405           & 1386.082                   & 0.407 & 21741 & $\{0,\pm 1\}$     \\ \hline
G30  & 2000 & 19990 & 100            & 3413           & 1396.632                    & 0.409 & 22292 & $\{0,\pm 1\}$     \\ \hline
G31  & 2000 & 19990 & -80            & 3310           & 1305.934                   & 0.395 & 22460 & $\{0,\pm 1\}$     \\ \hline
G59  & 5000 & 29570 & 96             & 6086           & 2274.439                   & 0.374 & 39462 & $\{0,\pm 1\}$     \\ \hline
G61  & 7000 & 17148 & 362            & 5796           & 2476.214                   & 0.427 & 14794 & $\{0,\pm 1\}$     \\ \hline
G64  & 7000 & 41459 & 527            & 8751           & 3381.164                   & 0.386 & 57647 & $\{0,\pm 1\}$     \\ \hline
\end{tabular}
\caption{Computational results of \QAOA $p=1$ expectation values in cut values for a subset of the G-set instances. Column \emph{Sum of weights} shows the sum of all $J_{ij}$ values. Column \emph{Best-known cut} shows the best classical results for the cut value, column \emph{\QAOA expectation cut} gives the best \QAOA expectation value. The best-known classical results were obtained from \cite{toshiba_benchmarking}. The column \emph{Ratio cut} shows the corresponding ratio of the \QAOA value versus the best classical result for the cut value. \emph{Time (s)} column shows the time required to obtain the full landscape and optimal angles, in seconds, on a \emph{c5n.large} AWS instance, which uses a \SI{3.0}{GHz} Intel Xeon Platinum Processor. \cite{c5n}}
\label{table:G-set_cut}
\end{table}

\begin{table}[t]
\begin{tabular}{c| S[table-format=6.0] S[table-format=6.0] S[table-format=+3.0] S[table-format=5.3] S[table-format=6.3] S[table-format=1.3] S[table-format=6.0] c}
Name & {$|V|$} & {$|E|$} & {Sum of weights} & {\QAOA expectation} & {$1.18\sqrt{|V| \times (|E|+|V|)}$} & {Ratio exp.} & {Time (s)} & {Weights} \\ \hline \hline
I10  & 800  & 19176 & -144           & 1706.929          & 4717.167          & 0.362 & 31376 & $\{0,\pm 1\}$       \\ \hline
I20  & 800  & 4672  & -128           & 787.059          & 2468.88           & 0.319 & 5269 & $\{0,\pm 1\}$        \\ \hline
I30  & 2000 & 19990 & 86             & 2820.98         & 7825.455          & 0.36 & 22648 & $\{0,\pm 1\}$       \\ \hline
I59  & 5000 & 29570 & 156            & 4940.085         & 15513.747         & 0.318 & 40740 & $\{0,\pm 1\}$       \\ \hline
I64  & 7000 & 41459 & 661            & 6914.15         & 21732.929         & 0.318 & 59894 & $\{0,\pm 1\}$       \\ \hline
I99 & 70000 & 140000 & -62 & 54809.093 & 143067.397 & 0.383 & 125363 & $\{0,\pm 1\}$ \\ \hline
I100 & 100000 & 150000 & 956 & 72847.211 & 186574.382 & 0.39 & 137774 & $\{0,\pm 1\}$ \\ \hline
\end{tabular}
\caption{Computational results of \QAOA $p=1$ expectation values in Ising energies for a subset of the G-set instances with external fields, the I-set. Plus two more custom generated instances, I99, and I100. Column \emph{Sum of weights} shows the sum of all $h_i$ and $J_{ij}$ values,  column \emph{\QAOA expectation} gives the best \QAOA expectation value. The column \emph{Ratio exp.} corresponds to the ratio of the \QAOA energy expectation value versus the optimal energy predicted by the  formula $1.18\sqrt{|V| \times (|E|+|V|)}$. \emph{Time (s)} column shows the time required to obtain the full landscape and optimal angles, in seconds, on an \emph{c5n.large} AWS instance, which uses a \SI{3.0}{GHz} Intel Xeon Platinum Processor. \cite{c5n}}
\label{table:G-set_ising_field}
\end{table}

{ \setlength{\tabcolsep}{10pt}
\begin{table}[t]
\begin{tabular}{c | S[table-format=4.0] S[table-format=5.0] S[table-format=2.2] S[table-format=2.1] 
S[table-format = 1.0e+1] S[table-format = 1.0e+1] 
S}
{Name} & {$|V|$} & {$|E|$}  & {$d_\textrm{ave}$} & {$\sigma_{d}$} & {$\beta_\textrm{dif}$} & {$\gamma_\textrm{dif}$} & {expectation dif.} \\ \hline\hline
G1   & 800  & 19176 & 47.94   & 6.3 & 3d-2  &  2d-4 &  \SI{0.3}{\percent}  \\ \hline
G2   & 800  & 19176 & 47.94   & 6.5 & 3d-2  &  2d-4 &  \SI{0.3}{\percent}      \\ \hline
G3   & 800  & 19176 & 47.94  & 6.7 & 3d-2  &  2d-4 &  \SI{0.3}{\percent}     \\ \hline
G4   & 800  & 19176 & 47.94   & 6.7 & 3d-2  &  2d-4 &  \SI{0.3}{\percent}      \\ \hline
G5   & 800  & 19176 & 47.94  & 6.6 & 3d-2  &  2d-4 &  \SI{0.3}{\percent}      \\ \hline
G6   & 800  & 19176 & 47.94   & 6.3 & 8d-4  &  2d-4 &  \SI{0.4}{\percent}       \\ \hline
G7   & 800  & 19176 & 47.94   & 6.5 & 8d-4  &  2d-4 &  \SI{0.4}{\percent}      \\ \hline
G8   & 800  & 19176 & 47.94  & 6.7 & 8d-4  &  2d-4 &  \SI{0.4}{\percent}      \\ \hline
G9   & 800  & 19176 & 47.94  & 6.7 & 8d-4  &  2d-4 &  \SI{0.4}{\percent}      \\ \hline
G10  & 800  & 19176 & 47.94   & 6.6 & 8d-4  &  2d-4 &  \SI{0.4}{\percent}     \\ \hline
G11  & 800  & 1600  & 4    & 0           & 8d-4  &  1d-2 & \SI{1}{\percent}     \\ \hline
G12  & 800  & 1600  & 4    & 0           & 8d-4  &  1d-2 & \SI{1}{\percent}     \\ \hline
G13  & 800  & 1600  & 4  & 0            & 8d-4  &  1d-2 & \SI{1}{\percent}    \\ \hline
G14  & 800  & 4694  & 11.74   & 11.5 & 5d-2  &  2d-3 & \SI{1.5}{\percent}     \\ \hline
G15  & 800  & 4661  & 11.65   & 12.1  & 5d-2  &  1d-3 & \SI{1.5}{\percent}     \\ \hline
G16  & 800  & 4672  & 11.68   & 11.6 & 5d-2  &  2d-3 & \SI{1.5}{\percent}     \\ \hline
G17  & 800  & 4667  & 11.67   & 11.8    & 5d-2  &  2d-3 & \SI{1.5}{\percent}     \\ \hline
G18  & 800  & 4694  & 11.74   & 11.5 & 8d-4  &  1d-3 & \SI{0.2}{\percent}  \\ \hline
G19  & 800  & 4661  & 11.65   & 12.1  & 8d-4  &  1d-3 & \SI{0.3}{\percent} \\ \hline
G20  & 800  & 4672  & 11.68   & 11.6 & 8d-4  &  1d-3 & \SI{0.3}{\percent}    \\ \hline
G21  & 800  & 4667  & 11.67   & 11.8    & 8d-4  &  2d-3 & \SI{0.3}{\percent}  \\ \hline
G27  & 2000 & 19990 & 19.99   & 4.5  & 8d-4  &  2d-3 & \SI{0.4}{\percent} \\ \hline
G28  & 2000 & 19990 & 19.99   & 4.4  & 8d-4  &  2d-3 & \SI{0.4}{\percent}  \\ \hline
G29  & 2000 & 19990 & 19.99   & 4.5  & 8d-4  &  2d-3 & \SI{0.4}{\percent}  \\ \hline
G30  & 2000 & 19990 & 19.99  &  4.5     & 8d-4  &  2d-3 & \SI{0.4}{\percent}  \\ \hline
G31  & 2000 & 19990 & 19.99   & 4.6   & 8d-4  &  2d-3 & \SI{0.4}{\percent}  \\ \hline
G59  & 5000 & 29570 & 11.83   & 15.8 & 8d-4  &  3d-3 & \SI{0.5}{\percent}  \\ \hline
G61  & 7000 & 17148 & 4.93   & 2.2     & 8d-4  &  2d-3 & \SI{0.3}{\percent}  \\ \hline
G64  & 7000 & 41459 & 11.85   & 16.3 & 8d-4  &  3d-3 & \SI{0.5}{\percent}  \\ \hline
\end{tabular}
\caption{Computational results of depth $p{=}1$ \QAOA optimal angles and expectation values in comparison to the universal angles\cite{farhi2019quantum} for a subset of the G-set instances. Column $d_\textrm{ave}$ shows the average degree of the graph. Column $\sigma_{d}$ the standard deviation of the degrees of the vertices. 
Columns $\beta_\textrm{dif}$ and $\gamma_\textrm{dif}$ show the difference in values of $\beta$ and $\gamma$, respectively, between optimal and universal angles. 
The column \emph{expectation dif.} shows the percentage difference between the expectation values obtained for optimal and universal angles. }
\label{table:G-set_optimal_angle_dif}
\end{table}}

\begin{figure}[t]
\centering
\includegraphics[width=0.4\textwidth]{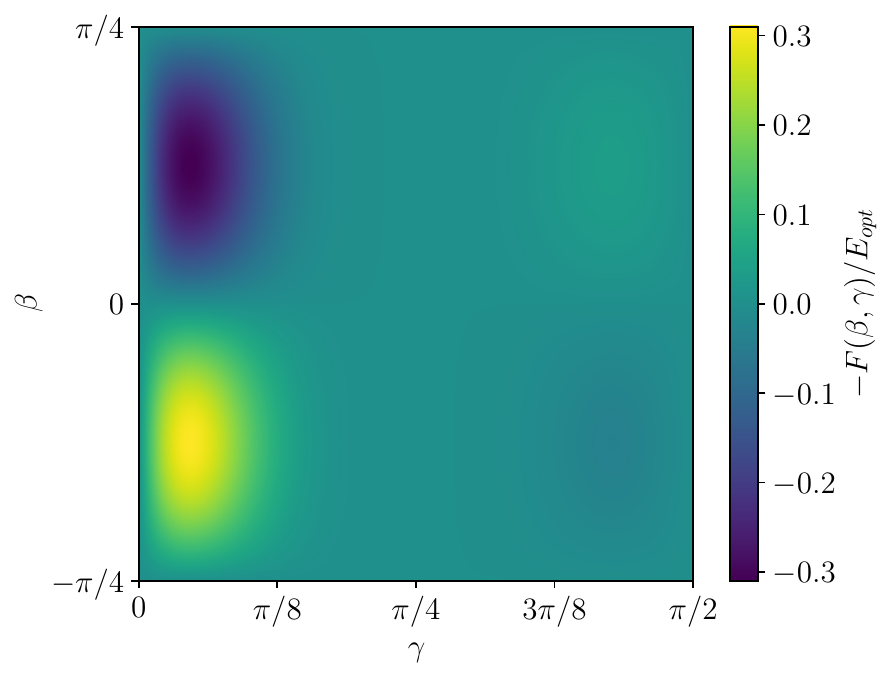}
\caption{\label{fig:G_set} Negative expectation-value landscape for the G64 instance with $\num{7000}$ vertices and $\num{41459}$ edges, obtained using the analytical formula Eq.~\ref{eq:main_res}. Normalized by the optimal energy $E_\textrm{opt}$.}
\end{figure}

\begin{figure}[t]
\centering
\includegraphics[width=0.6\textwidth]{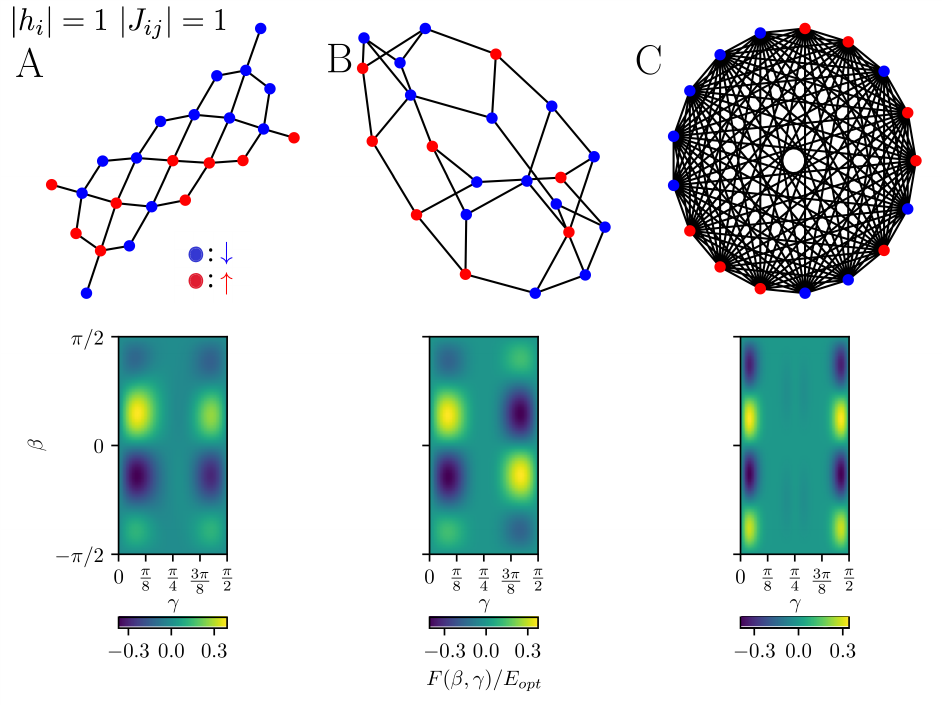}
\caption{\label{fig:google_plots_withfields} Expectation-value landscapes obtained for problem instances with $h_i, J_{ij} \in \{-1,1\}$, using the analytical formula Eq.~\ref{eq:main_res}. Normalized by the optimal energy $E_\textrm{opt}$. 
Problem graphs are shown above, with blue nodes representing those with $h_i=-1$, and red for those with $h_i=1$. 
(A) Sycamore grid problem with $23$ qubits. 
(B) $3$-regular graph with $22$ qubits. 
(C) Fully connected SK-model with $17$ qubits.}
\end{figure}

\begin{figure}[t]
\centering
\includegraphics[width=0.6\textwidth]{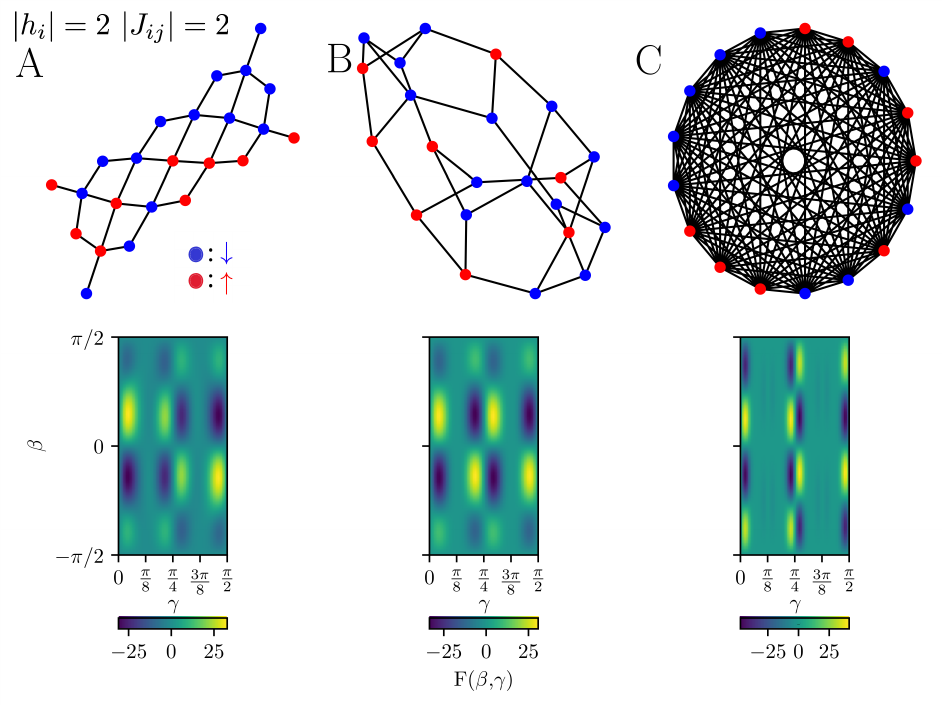}
\caption{\label{fig:google_plots_withfields2} Expectation-value landscapes obtained for problem instances with $h_i, J_{ij} \in \{-2,2\}$, using the analytical formula Eq.~\ref{eq:main_res}. 
Problem graphs are shown above, with blue nodes representing those with $h_i=-2$, and red for those with $h_i=2$. 
(A) Sycamore grid problem with $23$ qubits. 
(B) $3$-regular graph with $22$ qubits. 
(C) Fully connected SK-model with $17$ qubits.}
\end{figure}

\begin{figure}[t]
\centering
\includegraphics[width=0.6\textwidth]{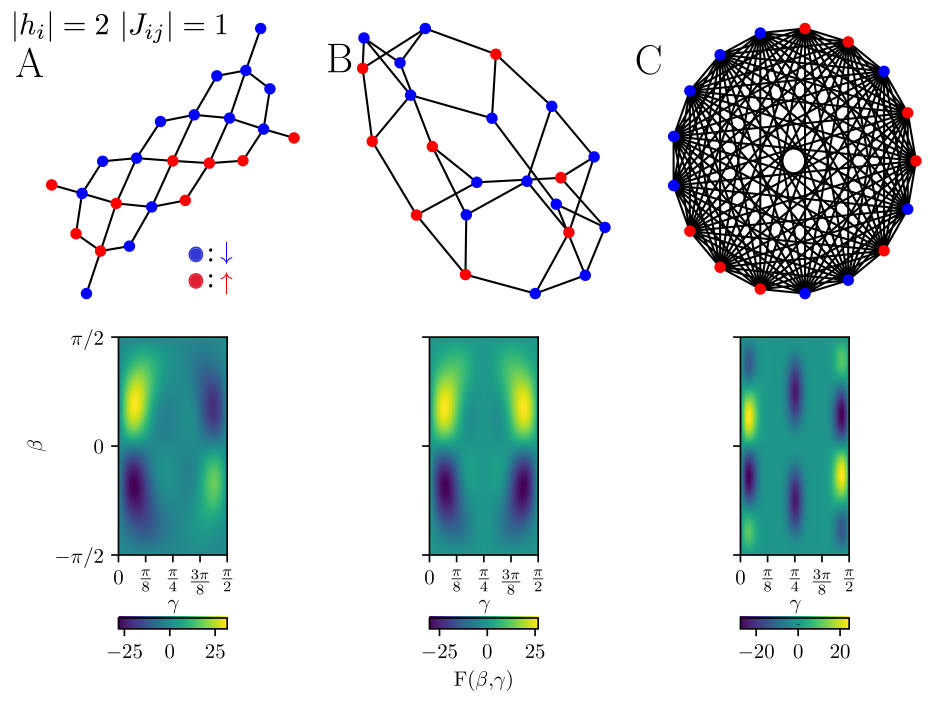}
\caption{\label{fig:google_plots_hgeJ} Expectation-value landscapes obtained for problem instances with $h_i \in \{-2,2\}$ and $ J_{ij} \in \{-1,1\}$, using the analytical formula Eq.~\ref{eq:main_res}.
Problem graphs are shown above, with blue nodes representing those with $h_i=-2$, and red for those with $h_i=2$. 
(A) Sycamore grid problem with $23$ qubits. 
(B) $3$-regular graph with $22$ qubits. 
(C) Fully connected SK-model with $17$ qubits.}
\end{figure}

\begin{figure}[t]
\centering
\includegraphics[width=0.6\textwidth]{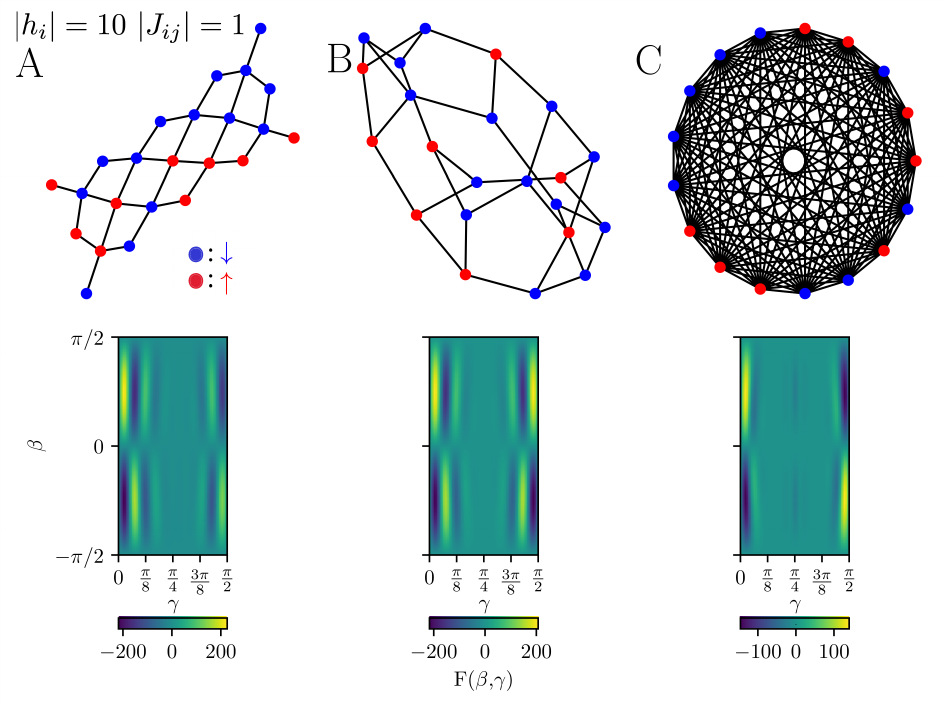}
\caption{\label{fig:google_plots_hgeJ10} Expectation-value landscapes obtained for problem instances with $h_i \in \{-10,10\}$ and $ J_{ij} \in \{-1,1\}$, using the analytical formula Eq.~\ref{eq:main_res}. 
Problem graphs are shown above, with blue nodes representing those with $h_i=-10$, and red for those with $h_i=10$. 
(A) Sycamore grid problem with $23$ qubits. 
(B) $3$-regular graph with $22$ qubits. 
(C) Fully connected SK-model with $17$ qubits.}
\end{figure}

\begin{figure}[t]
\centering
\includegraphics[width=0.6\textwidth]{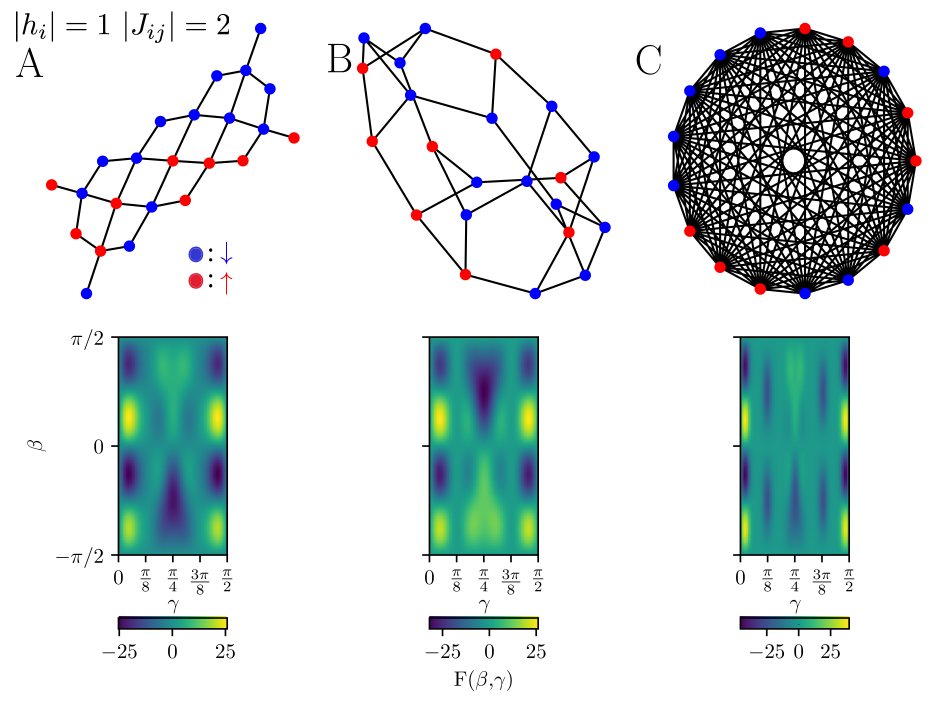}
\caption{\label{fig:google_plots_hleJ} Expectation-value landscapes obtained for problem instances with $h_i \in \{-1,1\}$ and $ J_{ij} \in \{-2,2\}$, using the analytical formula Eq.~\ref{eq:main_res}.
Problem graphs are shown above, with blue nodes representing those with $h_i=-1$, and red for those with $h_i=1$. 
(A) Sycamore grid problem with $23$ qubits. 
(B) $3$-regular graph with $22$ qubits. 
(C) Fully connected SK-model with $17$ qubits.}
\end{figure}

\begin{figure}[t]
\centering
\includegraphics[width=0.6\textwidth]{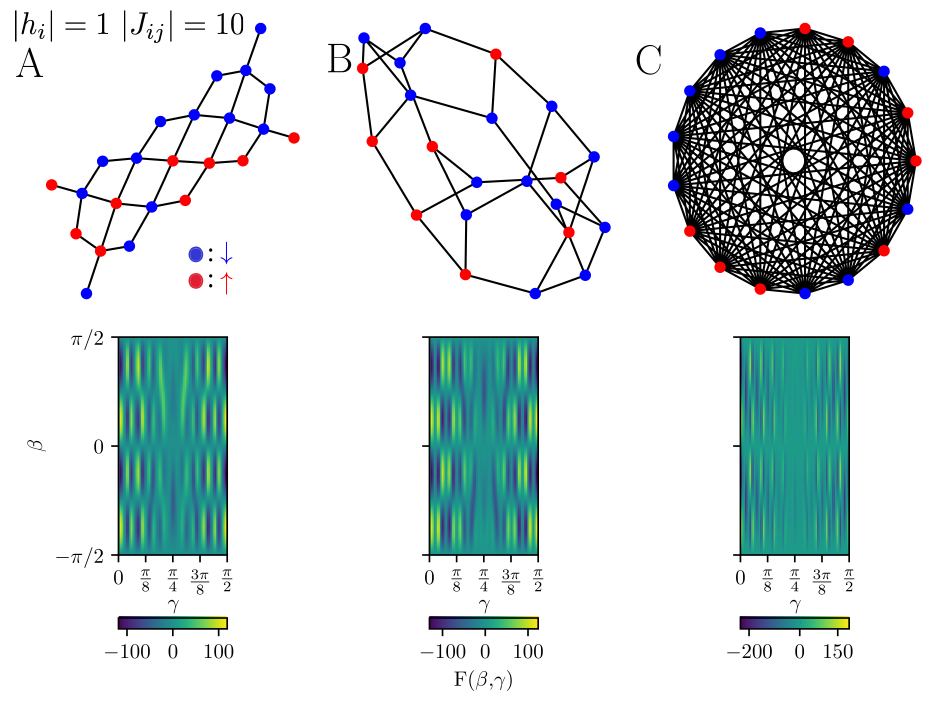}
\caption{\label{fig:google_plots_hleJ10} Expectation-value landscapes obtained for problem instances with $h_i \in \{-1,1\}$ and $ J_{ij} \in \{-10,10\}$, using the analytical formula Eq.~\ref{eq:main_res}.
Problem graphs are shown above, with blue nodes representing those with $h_i=-1$, and red for those with $h_i=1$. 
(A) Sycamore grid problem with $23$ qubits. 
(B) $3$-regular graph with $22$ qubits. 
(C) Fully connected SK-model with $17$ qubits.}
\end{figure}

\begin{figure}[t]
\centering
\includegraphics[width=0.4\textwidth]{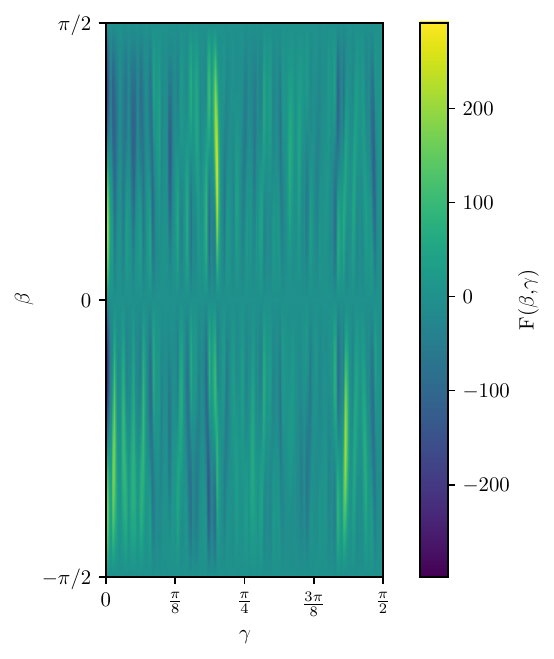}
\caption{\label{fig:google_plot_hJwild10} Expectation-value landscape, obtained using the analytical formula Eq.~\ref{eq:main_res}, for a 3-regular graph problem instance that has $16$ qubits, and $h_i \in \{-30, -29,\dots,29,30\}$ and $J_{ij} \in \{-30,30\}$.}
\end{figure}

\subsection{Numerical Results on Google \MaxCut and Sherrington-Kirkpatrick (SK) Instances}

The equations above allow us to replicate the expectation-value landscapes of the recent experimental work by Harrigan et al.\cite{google_qaoa}, which provides numerical support for the validity of our analytical results. 
In this section, we present our investigation of the problems studied in Ref.~\cite{google_qaoa}, as well as some similar but larger problem instances. The expectation value landscapes are obtained evaluating our analytical formula in a grid of $500 \times 500$ points, and from the results we obtain the optimal angles, and optimal expectation value. For clarity, Table~\ref{table:figure_summary} shows a summary of the parameters of the figures found in the text.

The three problems studied in \cite{google_qaoa} are different weighted \MaxCut instances with weights $-1$ and $+1$. 
The three specific instances used in \cite{google_qaoa} can be obtained from the \emph{ReCirq} repository~\cite{ReCirq}. 
We numerically evaluate Eq.~\ref{eq:main_res} for these instances, which gave us the energy-expectation landscapes for the single-layer \QAOA. 
The numerical evaluation of the expectations $F(\beta,\gamma)$ for varying values of both $\beta$ and $\gamma$ gives an energy-expectation landscapes, shown in Fig.~\ref{fig:google_plots}.
(Note that we plot $-F(\beta,\gamma)$ due to the normalization by a negative quantity of the expectation values in \cite{google_qaoa}.)
We observe that these landscapes, which we obtained using the analytical formula Eq.~\ref{eq:main_res}, are in excellent agreement with the brute-force-classical and quantum-hardware evaluations of the landscapes shown in \cite{google_qaoa}. 

In addition, the analytical formulas allow us to explore the landscapes of problem instances far larger than those possible to study on current experimental hardware. 
We have computed the landscapes for two $100$-spin (i.e., $100$-qubit) Ising problem instances: one instance has a $3$-regular graph structure, and the other is a Sherrington-Kirkpatrick instance. 
The graphs and their corresponding landscapes are shown in Fig.~\ref{fig:google_plots_morequbits}.

In relation with the previous section, note how the models shown in Fig.~\ref{fig:google_plots}.C and  Fig.~\ref{fig:google_plots_morequbits}.B correspond to the cases studied in Section~\ref{sec:SK_maxcut_expectation}. 
We predicted that $\sqrt{n-1} ~\gamma_\textrm{min} = 0.5$ for this bimodal distribution, and how that quantity will remain constant for any SK instances. 
For $n=11$ we obtain $\sqrt{10}~ \gamma^{(11)}_\textrm{min} \approx \sqrt{10} \cdot 0.158 \approx 0.5$, and for $n=100$, $\sqrt{99}~ \gamma^{(100)}_\textrm{min} \approx \sqrt{99} \cdot 0.0527 \approx 0.5$, confirming our theoretical prediction.

On the other hand, Fig.~\ref{fig:google_plots}.B and  Fig.~\ref{fig:google_plots_morequbits}.A correspond to the cases studied in Section~\ref{sec:reg_graphs_expectation}. Here we showed that $\sqrt{d} ~\gamma_\textrm{min} = 0.5$ for a bimodal distribution and $d$ regular graphs. In the first case, with $n=14$, we obtain $\sqrt{3}~\gamma^{(14)}_\textrm{min}\approx \sqrt{3} \cdot 0.316 \approx 0.5$. While for $n=100$, the equality reads $\sqrt{3}~\gamma^{(100)}_\textrm{min}\approx \sqrt{3} \cdot 0.306 \approx 0.5$.

We will now comment briefly on the features that one can observe in the landscapes shown in Figs.~\ref{fig:google_plots} and \ref{fig:google_plots_morequbits}. 
In Fig.~\ref{fig:google_plots}, the peaks are clearly narrower and closer to the edge for the $11$-spin SK instance than for the two ($23$-spin and $14$-spin) sparsely-connected instances. 
Furthermore, for the SK instance, additional peaks appear at $\gamma=\pi/4$. These peaks will vary in height depending on the distribution of the $\pm 1$ values in the graph.

In Fig.~\ref{fig:google_plots_morequbits}, we see that the landscape for the sparsely-connected ($3$-regular) instance with $100$ spins is the same as the landscape for the $3$-regular instance with $14$ spins---i.e., as the number of spins has increased, the landscape has not changed. In contrast, the landscape for the $100$-spin SK instance is qualitatively different than the landscape for the $11$-spin SK instance: the peaks in the landscape for the $100$-spin instance are narrower and closer to the edges of the $\gamma$ range, and the peaks at $\gamma=\pi/4$ for the $11$-spin instance are not present for the $100$-spin instance. The peaks at $\gamma=\pi/4$ also become less pronounced as we increase the size of the SK graph. From the results derived in theory Sections~\ref{sec:SK_maxcut_expectation} and \ref{sec:reg_graphs_expectation}, we showed above that $\gamma_\textrm{min}$ will remain constant for $3$-regular graphs, while decreasing with the number of spins for SK instances. The landscapes that we observe are a direct effect of this phenomenon. 

The couplings of an SK instance with $n$ spins is an $n$-regular graph. 
We can understand the narrowing and shifting of the peaks between the $n=11$ and $n=100$ SK instances we studied as an example of a more general phenomenon for regular graphs: the larger the degree of a regular graph, the narrower the peaks in the landscape will be. 
This phenomenon is explored in Sec.~\ref{sec:analytical_landscapes} and it has an important experimental consequence: the peak sharpness places a constraint on how accurately the angle $\gamma$ needs to be implemented in hardware, and so the necessary precision for specifying $\gamma$ will increase with the \emph{degree} of a regular graph.
 
While the expectation-value landscapes presented in Fig.~\ref{fig:google_plots} already show some qualitative differences from instance to instance, we will show later in this paper that landscapes can vary far more dramatically depending on the parameters of the cost Hamiltonian. 
Furthermore, we will study how the optimal $\gamma$ value, $\gamma_\textrm{min}$, that minimizes the expectation value, changes with those parameters. 
So far we have seen evidence that increased degree of regular graphs can cause $\gamma_\textrm{min}$ to be closer to $0$ or $\pi/2$, but we will find that the value of $\gamma_\textrm{min}$ can have a rather complex dependence on the cost-Hamiltonian parameters.

One of the potential benefits of using an analytical formula to determine the optimal angles instead of performing a brute-force search is speed, since the former requires on classical-computing calculations whereas the latter uses quantum simulations or runs on quantum hardware. This potential benefit is seen already with the instances studied in this section. To illustrate this, we simulated the expectation-value landscape of fully connected graphs with $23$ qubits using \emph{Cirq}~\cite{Cirq}; these calculations took approximately $1$ minute to complete. 
In comparison, using the analytical formula we could compute the landscape for $100$-qubit problems in approximately $4$ seconds. 
Both calculations were performed on a \emph{c5n.large} AWS instance. \cite{c5n}

\subsection{Numerical Results on G-Set \MaxCut Instances and Ising Instances with External Fields}

The G-set \cite{G_set} collection of benchmark instances for \MaxCut is widely used in the comparison of classical \MaxCut heuristics \cite{Benlic_2013,leleu2019destabilization,toshiba_benchmarking}. We chose to study single-layer QAOA's performance on G-set \MaxCut instances primarily because this enables comparison of QAOA's predicted performance against a wide range of heuristics, including the best-known classical methods. We also studied QAOA on Ising instances having non-zero external-field terms. We generated these instances for this study (details provided later); this was motivated by the fact that we are unaware of a standard benchmark set for such Ising instances.

We calculated the optimal \QAOA expectation values for a subset of the G-set instances having $\num{800}$, $\num{2000}$, $\num{5000}$, or $\num{7000}$ vertices and up to $\num{41459}$ edges. 
The results are divided into two tables; Table~\ref{table:G-set_Ising}, where results are expressed in terms of Ising energies, and Table~\ref{table:G-set_cut}, where we show them as cut values. The instances have weights that are drawn from two possible sets, depending on the instance: $\{0,1\}$ and $\{-1,0,+1\}$. For Table~\ref{table:G-set_Ising}, we derive in the appendix an analytical formula that allows us to estimate the optimal energy of a G-set instance, $E_\textrm{opt} \approx 1.18 \sqrt{|E|\times |V|}$. The ratio between the theoretical maximum and the \QAOA expectation, $r_\textrm{exp.}=F_\textrm{max}/E_\textrm{opt}$, ranges between $0.252-0.389$. This ratio is similar to the formula of the approximation ratio, $r=F_\textrm{max}/C_\textrm{max}$, that is frequently used in the literature to determine the efficiency of the \QAOA~\cite{farhi2014quantum,google_qaoa, wang2018quantum}. 
We just substitute the real maximum of the cost function by our theoretical approximation. 
The ratio for the best-known classical solution to the \QAOA expectation, $r_\textrm{Ising}=F_\textrm{max}/E^{cl}_\textrm{max}$, ranges between $0.36-0.475$. Note how the theoretical maximum is close to the classical best-known solution for most instances except G14-G17. For this set the \emph{Ratio Ising} and \emph{Ratio cut} are also particularly high. This leads us to believe that the best-known classical solution is not as good for these particular instances. This same logic applies to other instances with high values of \emph{Ratio Ising} e.g.\ G11. Additionally, the \QAOA expectation value is expected to be higher for instances with small amount of edges and common neighbours between vertices. These ratios can be used as a method to quickly determine the \QAOA expectation of a given instance; one can use the analytical formula to determine the theoretical maximum, and use the ratio range to determine the \QAOA expectation. The expectation value landscapes are very similar for the G-set instances, with similar optimal angles as predicted in \cite{brandao2018fixed}. Fig.~\ref{fig:G_set} shows the landscape of instance G64. 

We also calculated the best \QAOA expectation values for the $7$ Ising instances listed in Table~\ref{table:G-set_ising_field}. 
The first five of these instances were generated by adding weight $\{-1,+1\}$ external fields at random to instances of the G-set benchmark instances. 
We denote these instances as I$x$ where $x$ is the G instance label on which it is based. 
Additionally, two more instances, I99 and I100, were generated by assigning $\{-1,+1\}$ weights and fields at random.
The number of vertices and edges goes up to $\num{100000}$ and $\num{150000}$, respectively, in I100. 
An approximation for the optimal Ising energy of these instances was obtained using the analytical formula $E_\textrm{opt} \approx 1.18 \sqrt{|V| \times (|E|+|V|)}$, that is derived in the appendix. 
We obtained a ratio $r_\textrm{exp.}$ that ranges between $~0.318-0.39$ when comparing the theoretical optima and the \QAOA expectation. 
The values of the $r_\textrm{exp.}$ ratios are very similar to those obtained for \MaxCut instances. 
The \emph{Time} column displays in seconds the time required to obtain the full expectation value landscape and optimal angles.

In Table~\ref{table:G-set_optimal_angle_dif} we compare the optimal angles for a subset of the G-set instances with the universal angles suggested in \cite{farhi2019quantum},  which are only strictly valid for regular graphs.
To approximate this, we calculate the average degree $d_{\textrm{ave}}$ of the graph, and obtain the universal angles using this metric. 
We obtain the universal angles $\beta_\textrm{uni}, \gamma_\textrm{uni}$ using the formulas $\beta_\textrm{uni}=-\pi/8$ and $\gamma_\textrm{uni}=1/(2\sqrt{d_\textrm{ave}})$ derived in \cite{farhi2019quantum}. 
The true optimal angles for each instance of the G-set are obtained using a brute force approach such that the optimality of the angles is guaranteed up to the resolution of the search. 
The values displayed in columns $\beta_\textrm{dif}$ and $\gamma_\textrm{dif}$ correspond to the absolute difference between these two of angles. 
The standard deviation of the degrees of the vertices is captured by the standard expression $\sigma_d=(\sum_{i \in V} (d^2_i - d^2_{\textrm{ave}})/|V|)^{1/2}$. 
The $\sigma_d$ of an instance graph allows us to quantify how the graph is to a regular graph, which is defined by having zero deviation. 
Furthermore, we compare the difference in expectation values in the last column \emph{expectation dif.} by calculating the percentage difference using the formula, $\% = 200|F_\textrm{opti}-F_\textrm{uni}|/(F_\textrm{opti}+F_\textrm{uni})$, where $F_\textrm{opti}$ and $F_\textrm{uni}$ are the expectation values of the optimal and universal angles respectively. 

Overall we observe that the approach of using the average degree as a proxy for a regular degree is very successful, and that the obtained universal angles are similar to the optimal angles of the individual instances. With an observed maximum value of the \emph{expectation dif.} of \SI{1.5}{\percent} for all the instances with most showing values smaller or equal to \SI{1}{\percent}.
The cases with graphs closer to regular graphs do not show smaller differences in angles and expectations. 
How close the instance is to being regular is quantified by $\sigma_d$ as regular graphs are defined by $\sigma_d=0$.
Somewhat surprisingly, even for highly non-regular graphs (like G14--G21, G59, G64) the universal angles for regular graphs still work really well. 

In general, higher $d_\textrm{ave}$ values will cause the differences in angles to affect more strongly the difference in expectation value because the area with relevant expectation values in the landscape will be narrower as we increase $d_\textrm{ave}$. 
However, we do not observe this phenomena in our results.
Perturbations in $\gamma$ angles will also influence the difference in expectation value in a more relevant way than $\beta$ perturbations, due to the shape of the relevant expectation value area being longer in the $\beta$ direction. 
We also observe that having instances with weights $\{0,\pm 1\}$ or $\{0,1\}$ does not affect the expectation value difference in a relevant way. 
The described approach can be applied to any other problem instances either as a good starting point for the expectation value optimization or as a set of good angles to use in the \QAOA without the need for optimization.

\subsection{Numerical and Analytical Results on Ising Problem Expectation Value Landscapes}
\label{sec:analytical_landscapes}

In this section we will show the expectation-value landscapes of simple Ising problem instances with different weights, and derive analytic expressions that describe some of the observed phenomena.

First, we add external fields and study the case in which $h_i$ and $J_{ij}$ take only the values $\{-h,h\}$ assigned at random, where $h$ can take any (real) value, and the spin connectivity is that of a $d$-regular graph. 
For this particular case we can approximate $\gamma_\textrm{min}$ as
\begin{align}
\gamma_\textrm{min} & \approx \frac{1}{2h}\arctan{\frac{1}{\sqrt{d}}},\frac{1}{2h}( \pi - \arctan{\frac{1}{\sqrt{d}}}) ~.
\end{align}
We previously observed how the regularity of the graph affects the $\gamma_\textrm{min}$ value, and now we see that increasing the $h$ value will also bring $\gamma_\textrm{min}$ closer to the edges ($\gamma = 0$ and $\gamma = \pi/2$). The $h$ value also affects the $\gamma$-periodicity of the landscape: the period of the oscillations in the landscape as a function of $\gamma$ is proportional to $1/h$. For example, peaks that for $h=1$ were located at $\gamma=\pi/2$, for $h=2$ will be located at $\gamma=\pi/4$. The expectation value of $\gamma_\textrm{min}$ for the single-node contribution in Eq.~\ref{eq:main_res} reads,
\begin{align}
\langle C_{i} \rangle(\beta_\textrm{min}, \gamma_\textrm{min}) & =\frac{\sqrt{d}}{d+1}(1+\frac{1}{d})^{-(d-1)/2}.
\end{align}
This value decreases monotonically with $d$. Here we operate under the assumption that the single-node term will be the main contribution to the expectation value.
We observe the predicted phenomena in Fig.~\ref{fig:google_plots_withfields} and \ref{fig:google_plots_withfields2} where we plot the expectation landscapes for the cases $h=1$ and $h=2$, respectively. The first plot is, as expected, similar to what we encountered in the absence of external fields. Due to the presence of local minima and maxima with different expectation values, and additional features, the $\beta$ values of the plots now range between $-\pi/2$ and $\pi/2$. For $h=2$ we observe that the periodicity in $\gamma$ is divided by half, and there is a narrowing of the minimum and maximum peaks in $\gamma$. We had already seen (e.g., in Figures \ref{fig:google_plots} and \ref{fig:google_plots_morequbits}) that increasing number of edges per node causes the peaks in the landscape to become narrower in $\gamma$, and we now see that the peaks also become narrower when $h$ is increased.

The difference in expectation values for different minima is going to arise from the summation terms in Eq.~\ref{eq:main_res}. For the general case these terms have the form
\begin{equation}
[\cos(2 \gamma h)^o \cos(2 \gamma J)^p+\cos(2\gamma h)^q \cos(2 \gamma J)^r].   
\end{equation}
If by our choice of $o,p,q,r,h,J$ parameters we can make one term positive and another negative for the same value of $\gamma$, this will cause different expectation values for different minima.

Consider the case of problem instances with $h_i \in \{-rJ,rJ\}$ and $J_{ij} \in \{-J,J \}$ assigned at random, where the graph is again $d$-regular. In this case $J_{ij} \le h_i$ for all $i$ and $j$ (since $r$ is an integer). The following are two different approximations for $\gamma_\textrm{min}$ one can make depending on the value of $r$. For $r \le 5$,
\begin{align}
\gamma_\textrm{min} & \approx \frac{1}{2J}\arctan{(\frac{1}{\sqrt{d}}(1-0.15(r-1)))}.
\end{align}
For $r > 5$,
\begin{align}
\gamma_\textrm{min} & \approx \frac{1}{2r^{0.6}J}\arctan{(\frac{1}{\sqrt{d}})}.
\end{align}
These equations show that the relative increase of $h_i$ with regards to $J_{ij}$ is going to make $\gamma_\textrm{min}$ closer to the edges. This phenomenon is observed in the $r=2, 10$ cases shown in Fig.~\ref{fig:google_plots_hgeJ} and \ref{fig:google_plots_hgeJ10}, respectively. The $r=2$ case shows the peaks taking different shapes to what we previously observed. And for $r=10$, the generation of additional peaks at different locations in the landscape, and the narrowing of those. By evaluating the expectation value in $\gamma_\textrm{min}$ we observe that it decreases monotonically with $r$. Thus, an increase in $r$ will not only reduce $\gamma_\textrm{min}$ but it will generate additional peaks, the narrowing of those, a reduced optimal expectation value, and generate new shapes.

Alternatively, we can study the case $J_{ij} \ge h_i$, where we set $h_i \in \{ -h,h \}$ and $J_{ij} \in \{ -rh,rh \}$ assigned at random in a $d$ regular graph. For this case we can approximate $\gamma_\textrm{min}$,
\begin{align}
\gamma_\textrm{min} & \approx \frac{1}{2rh}\arctan{\frac{1}{\sqrt{d}}}
\end{align}
This equation shows how an increase in $r$ will again lead to $\gamma_\textrm{min}$ being closer to the edge. The expectation value of $\gamma_\textrm{min}$ for the single node contribution Eq.~\ref{eq:main_res} reads,
\begin{align}
\langle C_{i} \rangle(\beta_\textrm{min},\gamma_\textrm{min}) & =(1+\frac{1}{d})^{-\frac{d}{2}} \sin(\frac{\arctan(1/\sqrt{d)})}{r})~.
\end{align}
This expression decreases monotonically with $r$, as well as $d$, leading to a worse expectation value with increasing $r$.

Fig.~\ref{fig:google_plots_hleJ} and \ref{fig:google_plots_hleJ10} show the $r=2, 10$ cases respectively. The $r=2$ case shows additional peaks, narrowing of those, and new shapes. While for $r=10$ we have a complex landscape with a big number of narrow peaks with different shapes. Thus, an increase of $r$ for these kind of problem instances will lead to a $\gamma_\textrm{min}$ closer to the edge, a reduced optimal expectation value, additional peaks, narrowing of these, and new shapes.

All these phenomena are very similar to those observed in the $J_{ij} \le h_i$ case. However, unlike previously, the number of additional peaks is larger, the narrowing is much more noticeable, and the new shapes are easily observable. Thus, the increase of the relative $J_{ij}$ values has a bigger impact in generating wild expectation value landscapes. Notice also that the expectation value of some of these additional peaks is very similar to that of the optima. For the instance displayed in Fig.~\ref{fig:google_plots_hleJ10} the peaks are already very narrow and difficult to distinguish. This can cause difficulties when trying to set the optimal angles in an experimental setup due to precision issues. Our method can be useful in this case to identify the most suitable set of optimal angles that are experimentally more feasible.

We have shown that varying the parameters of the cost Hamiltonian the expectation value landscapes can look very different to the ones presented at the beginning of this paper; one example of this is shown in  Fig.~\ref{fig:google_plots_hleJ10}. This means that for certain Ising instances the computation of optimal angles is not straightforward and further study is required. As another example, consider the class of Ising instances where $h_i$ can take any integer value at random between $h$ and $-h$, i.e., $h_i  \in \{-h,\dots,h\}$, and $J_{ij} \in \{ -h, h \}$. The landscape of one such instance is shown in Fig.~\ref{fig:google_plot_hJwild10} for $h=30$. Here we no longer have any of the recognisable patterns that we had in the first landscapes, and the optimal angles are not at the edges.

More general and practically relevant problems may have complicated landscapes and require their own study. As we have shown, \QAOA landscapes can vary substantially depending on the class of problem instance, and finding the optimal angles using classical methods requires some care to be taken.

The objective of this paper is not to claim quantum advantage at $p=1$ but to present tools to investigate the expectation value landscapes present in \QAOA. Although our paper is limited to the study of a single layer, we see value in applying the findings presented here to higher $p$ versions of \QAOA. 

\section{Conclusions}

We introduced an analytical formula for the expectation value of single-layer ($p=1$) \QAOA circuits solving general Ising Hamiltonians, which allows for an efficient classical calculation of the optimal angles $\beta_\textrm{min}$, $\gamma_\textrm{min}$. 
We validated our analytical formula by numerically evaluating the \QAOA expectation values of several Ising problems studied in a recent experimental paper \cite{google_qaoa}. 
Furthermore, we predicted the energy landscapes for single-layer \QAOA with $100$ qubits, and benchmarked the expectation values for the G-set \MaxCut problems and custom Ising problems with hundreds to thousands of vertices/spins/qubits. 
For the G-set benchmark instances the ratio between the theoretical estimated maximum Ising energy and the \QAOA expectation for Ising energy was in the range $~0.252-0.389$; for the Ising problems with external fields that we tested, the ratio was in the range $~0.318-0.39$. 
In order to compute these ratios---since we do not know the true optimal energy for each of such large Ising instances---we derive an analytical formula that allows one to estimate the optimal energy of an Ising or \MaxCut instance with weights $\{-1,0,+1\}$. 
The largest instance that we estimated the accuracy of the single-layer \QAOA for had $\num{100000}$ vertices and $\num{150000}$ edges, and found that it achieved a ratio of $\approx 0.39$.

For the G-set \MaxCut instances, we also computed the ratio between the expected cut value from the single-layer \QAOA and the best-known cut value (obtained using classical heuristic methods) for each instance. 
For the G-set instances that are \emph{unweighted} (i.e., $J_{ij} \in \{0,1\}$), the ratios are all $\geq 0.85$---meaning that we can expect the \QAOA to find \MaxCut solutions for these instances that have at least $\SI{85}{\percent}$ of the maximum (known) cut value. 
The accuracy of the solutions one would obtain by running the single-layer \QAOA on such large instances in practice (under the assumption of noiseless quantum hardware) should be even higher, because each run of the \QAOA will yield a single sample from a distribution that has a \emph{mean} solution value with $\sim \SI{85}{\percent}$ accuracy, and so multiple runs of the \QAOA are likely to yield some samples that have solution values better than that of the mean value. Two related open questions of great interest are \emph{what the shape of the output-sample distribution from the \QAOA is} and \emph{what the probabilities of measuring solutions with energies much better than the mean energy are} \cite{pagano2020quantum}.

We also note that the multi-layer \QAOA cannot perform worse than the single-layer \QAOA, since one can always choose to set the angles for all layers after the first one to be equal to $0$, and hence have the layers after the first layer perform the identity operation, so our results provide a lower bound for the performance of the multi-layer \QAOA.

While the numerical evaluation of our analytical formula has polynomial scaling, the wall-clock times for our current classical code implementation are typically $\gg 1000$~seconds for Ising instances comprising $|V| \gg 100$ spins (exact times are given in Tables \ref{table:G-set_cut} and \ref{table:G-set_ising_field}). An area for future work is to increase the speed of the classical code for computing optimal angles.

Finally, we studied the effect that the values of the external-field terms $h_i$ and spin-spin-coupling terms $J_{ij}$ have on the expectation-value landscape and the optimal $\gamma$ parameter, $\gamma_\textrm{min}$. We showed that an increase of the external fields or spin-spin couplings appear, in general, to lead to the generation of additional peaks in the landscape, the narrowing of these peaks, new shapes, a reduction in optimal expectation value, and $\gamma_\textrm{min}$ being closer to the edges. A relative increase in $J_{ij}$ has a bigger impact in the generation of complex expectation landscapes. We found that a broad class of Ising Hamiltonians have highly complex landscapes that are very dissimilar to the ones that we are aware of in the literature. Our analytical methods for finding optimal angles may be particularly useful in these cases, especially since angle searches performed in situ using quantum hardware may easily ``miss'' very narrow features in the landscapes.

\emph{Note added in preparation:} Related work studying the single-layer \QAOA numerically appeared on the arXiv on 29 November 2020: \cite{bravyi2020hybrid}, which investigates the performance of the \QAOA on a particular subclass of MAX-CUT instances related to graph coloring.

\section{Acknowledgements}

The authors thank V.~Putz, J.\,I.~Adame, R.~Parrish and E.~Farhi for useful discussions, and
J.\,I.~Adame for a thorough reading of a draft of this paper. PLM acknowledges membership of the CIFAR Quantum Information Science Program as an Azrieli Global Scholar.

\section{Data availability}

The I-set dataset, which comprises Ising instances generated from a subset of the G-set \cite{G_set} instances by adding external fields, and two additional large Ising instances unrelated to G-set, is available at \href{https://doi.org/10.5281/zenodo.5258011}{https://doi.org/10.5281/zenodo.5258011}.

\bibliographystyle{ieeetr}
\bibliography{references}

\begin{thebibliography}{10}

\bibitem{farhi2014quantum}
E.~Farhi, J.~Goldstone, and S.~Gutmann, ``A quantum approximate optimization
  algorithm,'' {\em arXiv:1411.4028}, 2014.

\bibitem{moll2018quantum}
N.~Moll, P.~Barkoutsos, L.~S. Bishop, J.~M. Chow, A.~Cross, D.~J. Egger,
  S.~Filipp, A.~Fuhrer, J.~M. Gambetta, M.~Ganzhorn, {\em et~al.}, ``Quantum
  optimization using variational algorithms on near-term quantum devices,''
  {\em Quantum Science and Technology}, vol.~3, no.~3, p.~030503, 2018.

\bibitem{crooks2018performance}
G.~E. Crooks, ``Performance of the quantum approximate optimization algorithm
  on the maximum cut problem,'' {\em arXiv:1811.08419}, 2018.

\bibitem{zhou2018quantum}
L.~Zhou, S.-T. Wang, S.~Choi, H.~Pichler, and M.~D. Lukin, ``Quantum
  approximate optimization algorithm: performance, mechanism, and
  implementation on near-term devices,'' {\em Physical Review X}, vol.~10,
  no.~2, p.~021067, 2020.

\bibitem{hogg2000quantum}
T.~Hogg and D.~Portnov, ``Quantum optimization,'' {\em Information Sciences},
  vol.~128, no.~3-4, pp.~181--197, 2000.

\bibitem{trugenberger2002quantum}
C.~A. Trugenberger, ``Quantum optimization for combinatorial searches,'' {\em
  New Journal of Physics}, vol.~4, no.~1, p.~26, 2002.

\bibitem{lucas2014ising}
A.~Lucas, ``Ising formulations of many {NP} problems,'' {\em Frontiers in
  Physics}, vol.~2, p.~5, 2014.

\bibitem{google_qaoa}
M.~P. Harrigan, K.~J. Sung, M.~Neeley, K.~J. Satzinger, F.~Arute, K.~Arya,
  J.~Atalaya, J.~C. Bardin, R.~Barends, S.~Boixo, M.~Broughton, B.~B. Buckley,
  D.~A. Buell, B.~Burkett, N.~Bushnell, Y.~Chen, Z.~Chen, B.~Chiaro,
  R.~Collins, W.~Courtney, S.~Demura, A.~Dunsworth, D.~Eppens, A.~Fowler,
  B.~Foxen, C.~Gidney, M.~Giustina, R.~Graff, S.~Habegger, A.~Ho, S.~Hong,
  T.~Huang, L.~B. Ioffe, S.~V. Isakov, E.~Jeffrey, Z.~Jiang, C.~Jones,
  D.~Kafri, K.~Kechedzhi, J.~Kelly, S.~Kim, P.~V. Klimov, A.~N. Korotkov,
  F.~Kostritsa, D.~Landhuis, P.~Laptev, M.~Lindmark, M.~Leib, O.~Martin, J.~M.
  Martinis, J.~R. McClean, M.~McEwen, A.~Megrant, X.~Mi, M.~Mohseni,
  W.~Mruczkiewicz, J.~Mutus, O.~Naaman, C.~Neill, F.~Neukart, M.~Y. Niu, T.~E.
  O'Brien, B.~O'Gorman, E.~Ostby, A.~Petukhov, H.~Putterman, C.~Quintana,
  P.~Roushan, N.~C. Rubin, D.~Sank, A.~Skolik, V.~Smelyanskiy, D.~Strain,
  M.~Streif, M.~Szalay, A.~Vainsencher, T.~White, Z.~J. Yao, P.~Yeh,
  A.~Zalcman, L.~Zhou, H.~Neven, D.~Bacon, E.~Lucero, E.~Farhi, and R.~Babbush,
  ``Quantum approximate optimization of non-planar graph problems on a planar
  superconducting processor,'' {\em Nature Physics}, vol.~17, no.~3,
  pp.~332--336, 2021.

\bibitem{streif2020training}
M.~Streif and M.~Leib, ``Training the quantum approximate optimization
  algorithm without access to a quantum processing unit,'' {\em Quantum Science
  and Technology}, vol.~5, no.~3, p.~034008, 2020.

\bibitem{G_set}
Y.~Ye, ``Gset test problems,'' 2003.
\newblock
  \href{https://web.stanford.edu/~yyye/yyye/Gset/}{https://web.stanford.edu/~yyye/yyye/Gset/}.

\bibitem{wang2018quantum}
Z.~Wang, S.~Hadfield, Z.~Jiang, and E.~G. Rieffel, ``Quantum approximate
  optimization algorithm for {MaxCut}: A fermionic view,'' {\em Physical Review
  A}, vol.~97, no.~2, p.~022304, 2018.
\newblock arXiv:1706.02998.

\bibitem{Bravyi2019ObstaclesTS}
S.~Bravyi, A.~Kliesch, R.~Koenig, and E.~Tang, ``Obstacles to variational
  quantum optimization from symmetry protection,'' {\em Physical Review
  Letters}, vol.~125, no.~26, p.~260505, 2020.

\bibitem{Barahona_1982}
F.~Barahona, ``On the computational complexity of {Ising} spin glass models,''
  {\em Journal of Physics A: Mathematical and General}, vol.~15,
  pp.~3241--3253, Oct 1982.

\bibitem{farhi2019quantum}
E.~Farhi, J.~Goldstone, S.~Gutmann, and L.~Zhou, ``The quantum approximate
  optimization algorithm and the {Sherrington-Kirkpatrick} model at infinite
  size,'' {\em arXiv preprint arXiv:1910.08187}, 2019.

\bibitem{toshiba_benchmarking}
Y.~Matsuda, ``Benchmarking the {MAX-CUT} problem on the simulated bifurcation
  machine,'' {\em Medium}, 2019.
\newblock
  \url{https://medium.com/toshiba-sbm/benchmarking-the-max-cut-problem-on-the-simulated-bifurcation-machine-e26e1127c0b0}.

\bibitem{c5n}
``Amazon ec2 instance types.''
\newblock \url{https://aws.amazon.com/ec2/instance-types/}.

\bibitem{ReCirq}
{Google Quantum AI Team and collaborators}, ``Recirq,'' Oct. 2020.
\newblock \url{https://doi.org/10.5281/zenodo.4091470}.

\bibitem{Cirq}
{Google Quantum AI Team and collaborators}, ``Cirq,'' Oct. 2020.
\newblock \url{https://doi.org/10.5281/zenodo.4062499}.

\bibitem{Benlic_2013}
U.~Benlic and J.-K. Hao, ``Breakout local search for the {Max-Cut} problem,''
  {\em Engineering Applications of Artificial Intelligence}, vol.~26,
  p.~1162–1173, 03 2013.

\bibitem{leleu2019destabilization}
T.~Leleu, Y.~Yamamoto, P.~L. McMahon, and K.~Aihara, ``Destabilization of local
  minima in analog spin systems by correction of amplitude heterogeneity,''
  {\em Physical Review Letters}, vol.~122, no.~4, p.~040607, 2019.

\bibitem{brandao2018fixed}
F.~G. S.~L. Brand{\~a}o, M.~Broughton, E.~Farhi, S.~Gutmann, and H.~Neven,
  ``For fixed control parameters the quantum approximate optimization
  algorithm's objective function value concentrates for typical instances,''
  {\em arXiv:1812.04170}, 2018.

\bibitem{pagano2020quantum}
G.~Pagano, A.~Bapat, P.~Becker, K.~S. Collins, A.~De, P.~W. Hess, H.~B. Kaplan,
  A.~Kyprianidis, W.~L. Tan, C.~Baldwin, {\em et~al.}, ``Quantum approximate
  optimization of the long-range {I}sing model with a trapped-ion quantum
  simulator,'' {\em Proceedings of the National Academy of Sciences}, vol.~117,
  no.~41, pp.~25396--25401, 2020.

\bibitem{bravyi2020hybrid}
S.~Bravyi, A.~Kliesch, R.~Koenig, and E.~Tang, ``Hybrid quantum-classical
  algorithms for approximate graph coloring,'' {\em arXiv:2011.13420}, 2020.

\bibitem{montanari2019optimization}
A.~Montanari, ``Optimization of the {S}herrington-{K}irkpatrick
  {H}amiltonian,'' in {\em 2019 IEEE 60th Annual Symposium on Foundations of
  Computer Science (FOCS)}, pp.~1417--1433, IEEE, 2019.

\bibitem{parisi1979infinite}
G.~Parisi, ``Infinite number of order parameters for spin-glasses,'' {\em
  Physical Review Letters}, vol.~43, no.~23, p.~1754, 1979.

\end{thebibliography}

\appendix
\section{Derivation of Analytical Expression for General Expectation}
\label{app:traces}

Given an undirected graph $G = (V, E)$ with vertices $V=\{1,\dots,n\}$ and edges $E\subset V\times V$, combined with external fields $h_i$ for the vertices, and coupling strengths $J_{ij}$ on the edges, we aim to find the configuration $s\in\{-1,+1\}^n$ that minimizes the cost function
\begin{align}
C & =   \sum_{i\in V} h_i s_i  + \sum_{(i,j)\in E} J_{ij} s_i s_j. 
\end{align}
This cost function is equivalent to the cost Hamiltonian
\begin{align}
H_C &= \sum_{i\in V} C_i + \sum_{(i,j)\in E} C_{ij} , 
\end{align}
where 
\begin{align}
\left\{
\begin{array}{rl} 
C_{i} & = h_i \sigma^z_i\\
C_{ij} & = J_{ij} \sigma^z_i \sigma^z_j
\end{array}
\right.
\end{align}
The unitary transformation $U$ for a \QAOA circuit with depth $p=1$ is defined as
\begin{align}
U & = U_B^\beta U_C^\gamma  
\end{align}
where
\begin{align}
U_C^\gamma & = e^{-\imag\gamma H_C}= e^{-\imag\gamma (\sum_{ij} J_{ij} \sigma^z_i \sigma^z_j + \sum h_i \sigma^z_i)} \\
& = \prod_{(i,j)\in E} (\cos(J_{ij} \gamma)\sigma_i^0\sigma_j^0 - \imag \sin( J_{ij}\gamma) \sigma^z_i \sigma^z_j) \prod_{i\in V} (\cos(h_i \gamma)\sigma_i^0 - \imag \sin(h_i\gamma) \sigma^z_i) \\
U_B^\beta & = e^{-\imag\beta H_B}= e^{-\imag\beta \sum_i \sigma^x_i} \\
& = \prod_{i\in V} ( \cos(\beta)\sigma_i^0 - \imag \sin(\beta) \sigma^x_i )  .
\end{align}
Applying $U = U_B^\beta U_C^\gamma$ to an initial, uniform superposition of all bit strings, $\ket{\psi_0}$, we obtain the final \QAOA state
\begin{align}
\ket{\beta,\gamma}& =U \ket{\psi_0}.
\end{align}
The expectation value of $H_C$ for this final state reads
\begin{align}
F(\beta, \gamma) & = \sum_{i\in V} \langle C_{i} \rangle  + \sum_{(i,j)\in E} \langle C_{ij} \rangle 
\end{align}
where
\begin{align}
\left\{
\begin{array}{rl}
\langle C_{i} \rangle  
& = \bra{\beta,\gamma}C_i\ket{\beta,\gamma} 
= \bra{\psi_0}U^\dagger C_i U \ket{\psi_0} 
= h_i \Tr [\rho_0 U^{\dagger} \sigma^z_{i} U]\\
\langle C_{ij} \rangle 
& = \bra{\beta,\gamma}C_{ij}\ket{\beta,\gamma} 
= \bra{\psi_0}U^\dagger C_{ij} U \ket{\psi_0} 
= J_{ij}\Tr [\rho_0 U^{\dagger} \sigma^z_i\sigma^z_j U]\\
\end{array}
\right.
\label{eq:app_trace_ci}
\end{align}
and $\rho_0 =\ket{\psi_0} \bra{\psi_0} $ is the initial density matrix. 

 Resolving the $U_B$ unitaries, the terms inside the traces read
\begin{align}
\left\{\begin{array}{rl}
 U^{\dagger}  \sigma^z_i U 
& = 
  U^{-\gamma}_C  (\cos(2\beta) \sigma^z_i + \sin(2\beta) \sigma^y_i) U_C^\gamma \\
 U^{\dagger}  \sigma^z_i \sigma^z_j U 
& = 
  U^{-\gamma}_C (\cos(2 \beta) \sigma^z_i + \sin(2 \beta) \sigma^y_i ) (\cos(2 \beta)\sigma^z_j+\sin(2 \beta) \sigma^y_j)  
U^\gamma_C
\end{array}
\right.
\end{align}

As a result, we can break down the computation into the evaluation of the following 6 traces, which we group according to the number of $\sigma^y$ terms:
\begin{align}
 &  \Tr[\rho_0 U^{-\gamma}_C \sigma^z_i U^\gamma_C],  \quad    \Tr[\rho_0 U^{-\gamma}_C \sigma^z_i\sigma_j^z U^\gamma_C],   \label{eq:ztraces}\\
 &   \Tr[\rho_0 U^{-\gamma}_C \sigma^y_i U^\gamma_C], \label{eq:ytraces} \\
 &  \Tr[\rho_0 U^{-\gamma}_C \sigma^y_i\sigma_j^z U^\gamma_C], \quad  \Tr[\rho_0 U^{-\gamma}_C \sigma^z_i\sigma_j^y U^\gamma_C], \label{eq:yztraces}  \\
 &  \Tr[\rho_0 U^{-\gamma}_C \sigma^y_i\sigma_j^y U^\gamma_C]. & \label{eq:yytraces}
\end{align}

Note that the terms in $U^{\pm\gamma}_C$ that do not involve the vertices $i$ and $j$ commute with $\sigma^*_i\sigma^*_j$, hence those terms will cancel each other. 
When evaluating $U^{-\gamma}_C\sigma^*_i\sigma^*_j U_C^\gamma$ we therefore only have to keep track of the terms of $C$ involving $h_i$, $h_j$, $J_{ij}$, $J_{ik}$, and $J_{jk}$ (with $k\neq i,j$). 
Also, because $\rho_0 = 2^{-n}\prod_{j=1}^n (\sigma^0_j+\sigma^x_j)$, the trace $\Tr[\rho_0 M]$ only depends on the $\{\sigma^0,\sigma^x\}^n$ components of $M$, that is: 
\begin{align}\label{eq:trace0x}
    \Tr\left[\rho_0 \sum_{s\in\{0,x,y,z\}^n} c_s \sigma_1^{s_1}\cdots \sigma_n^{s_n}\right] & = 
    \sum_{s\in\{0,x\}^n}c_s.
\end{align}

Going back to the six traces, because $\sigma^z_i$ and $\sigma^z_j$ commute with $U_C$ we see that the traces of Equation~\ref{eq:ztraces} equal $0$. 

To tackle the traces of Equation~\ref{eq:ytraces} we perform the calculation
\begin{align}
U^{-\gamma}_C   \sigma^y_i U_C^\gamma & = 
 \prod_{(i,k)\in E} (\cos(J_{ik} \gamma)\sigma_i^0\sigma_k^0 + \imag \sin( J_{ik}\gamma) \sigma^z_i \sigma^z_k) 
\cdot (\cos(h_i \gamma)\sigma_i^0 + \imag \sin(h_i\gamma) \sigma^z_i) \cdot \sigma^y_i \\ 
& \quad \times (\cos(h_i \gamma)\sigma_i^0 - \imag \sin(h_i\gamma) \sigma^z_i) 
\prod_{(i,k)\in E} (\cos(J_{ik} \gamma)\sigma_i^0\sigma_k^0 - \imag \sin( J_{ik}\gamma) \sigma^z_i \sigma^z_k) \\
& = \prod_{(i,k)\in E} (\cos(J_{ik} \gamma)\sigma_i^0\sigma_k^0 + \imag \sin( J_{ik}\gamma) \sigma^z_i \sigma^z_k) \cdot 
(\sin(2h_i\gamma) \sigma^x_i + \cos(2h_i \gamma)\sigma^y_i )\\
& \quad \times 
\prod_{(i,k)\in E} (\cos(J_{ik} \gamma)\sigma_i^0\sigma_k^0 - \imag \sin( J_{ik}\gamma) \sigma^z_i \sigma^z_k)
\\
& = (\sin(2h_i\gamma) \sigma^x_i\sigma^0_k + \cos(2h_i\gamma) \sigma^y_i\sigma^0_k) \prod_{(i,k)\in E}((\cos(J_{ik}\gamma))^2-(\sin(J_{ik}\gamma))^2) \\
& \quad + 2 (\cos(2h_i\gamma) \sigma^x_i\sigma^z_k - \sin(2h_i\gamma) \sigma^y_i\sigma^z_k) \prod_{(i,k)\in E}\cos(J_{ik}\gamma) \sin(J_{ik}\gamma)
\end{align}

Hence, there is only term that has a non-zero trace and we obtain the following result for Equation~\ref{eq:ytraces}:

\begin{align}
\Tr[\rho_0 U^{-\gamma}_C \sigma_i^y U^\gamma_C]& =  
\Tr
\left[\rho_0\sin(2h_i\gamma)\prod_{(i,k)\in E}((\cos(J_{ik}\gamma))^2-(\sin(J_{ik}\gamma))^2)\sigma^x_i\sigma^0_k\right] \\
& =  \sin(2 h_i \gamma ) \prod_{(i,k) \in E} \cos(2 J_{ik} \gamma).
\end{align}
For trace Expressions~\ref{eq:yztraces} we use the fact that $\rho_0 U_C^{-\gamma}\sigma_i^y\sigma_j^z U_C^\gamma = \rho_0 U_C^{-\gamma}\sigma_i^y U_C^\gamma \sigma_j^z$, and by some additional algebra similar to the calculation above we obtain

\begin{align}
\Tr[\rho_0 U_C^{-\gamma}\sigma_i^y U_C^\gamma \sigma_j^z] & = 
\Tr
\left[\rho_0 \cos(2h_i\gamma) 2 \sin(J_{ij} \gamma) \cos(J_{ij} \gamma)  \prod_{\substack{(i,k) \in E\\k \neq j}} \cos(2 J_{ik} \gamma) \sigma^x_i \sigma^z_j \sigma^0_k \cdot \sigma_j^z \right] \\
& =
\cos(2 h_i \gamma ) \sin(2 J_{ij} \gamma)   \prod_{\substack{(i,k) \in E\\k \neq j}} \cos(2 J_{ik} \gamma).
\end{align}

By symmetry, we have, $\rho_0 U_C^{-\gamma}\sigma_i^z\sigma_j^y U_C^\gamma = \rho_0  U_C^{-\gamma}\sigma_j^y U_C^\gamma \sigma_i^z $ and,

\begin{align}
    \Tr[\rho_0 \sigma_i^z U_C^{-\gamma}\sigma_j^y U_C^\gamma] & = 
      \cos(2 h_j \gamma ) \sin(2 J_{ij} \gamma)  \prod_{\substack{(j,k) \in E\\k \neq i}} \cos(2 J_{jk} \gamma).
\end{align}

We thus have to calculate the remaining case, Expression~\ref{eq:yytraces}. We used Mathematica to perform and verify the algebraic manipulations necessary to calculate Expression~\ref{eq:yytraces} and obtain the result of Eq.~\ref{eq:app_traceyy}, since a large number of terms is involved.

\begin{align}
    \Tr[\rho_0 U^{-\gamma}_C \sigma^y_i\sigma_j^y U_C^\gamma] & = 
      -\frac{1}{2}  \prod_{\substack{(ik)\in E\\ (jk)\notin E}} \cos(2 \gamma J_{ik})  \prod_{\substack{(jk)\in E\\ (ik)\notin E}}\cos(2\gamma J_{jk})  \\
& \qquad \times \Big[\cos(2 \gamma (h_i{+}h_j)) \prod_{\substack{(ik)\in E\\(jk)\in E}} \cos(2 \gamma (J_{ik}{+}J_{jk})) - 
\cos(2 \gamma (h_i{-}h_j)) \prod_{\substack{(ik)\in E\\(jk)\in E}} \cos(2 \gamma (J_{ik}{-}J_{jk}))\Big]
\label{eq:app_traceyy}
\end{align}

For this derivation the following trigonometric identities were used,

\begin{align}
\cos(2 \gamma (h_i{+}h_j)) \cos(2 \gamma (J_{ik}{+}J_{jk})) = \\
(\cos(2 \gamma h_i) \cos(2 \gamma h_j) - \sin(2 \gamma h_i) \sin(2 \gamma h_j))(\cos(2 \gamma J_{ik}) \cos(2 \gamma J_{jk}) - \sin(2 \gamma J_{ik}) \sin(2 \gamma J_{jk})) = \\
\cos(2 \gamma (h_i+h_j+J_{ik}+J_{jk}))+\cos(2 \gamma (h_i+h_j-J_{ik}-J_{jk}))
\end{align}

\begin{align}
\cos(2 \gamma h_i) = (\cos(\gamma h_i))^2-(\sin(\gamma h_i))^2
\end{align}

Resulting in the expression,

\begin{align} 
\left\{\begin{aligned}
\label{eq:main_res_app}
\langle C_i \rangle & = 
h_i \sin(2\beta)\sin(2\gamma h_i)\prod_{(i,k)\in E}\cos(2\gamma J_{ik}) \\
\begin{split}
\langle C_{ij} \rangle & = 
\frac{J_{ij} \sin(4\beta)}{2} \sin(2 \gamma J_{ij}) 
\Big[ \cos(2 \gamma h_i) \prod_{\substack{(ik) \in E\\ k \neq j}} \cos(2 \gamma J_{ik} ) + \cos(2 \gamma h_j) \prod_{\substack{(jk) \in E \\ k \neq i}} \cos(2 \gamma J_{jk})\Big]  \\
& \quad -\frac{J_{ij}}{2} (\sin(2\beta))^2  \prod_{\substack{(ik)\in E\\ (jk)\notin E}} \cos(2 \gamma J_{ik})  \prod_{\substack{(jk)\in E\\ (ik)\notin E}}\cos(2\gamma J_{jk})  \\
& \qquad \times \Big[\cos(2 \gamma (h_i{+}h_j)) \prod_{\substack{(ik)\in E\\(jk)\in E}} \cos(2 \gamma (J_{ik}{+}J_{jk})) - 
\cos(2 \gamma (h_i{-}h_j)) \prod_{\substack{(ik)\in E\\(jk)\in E}} \cos(2 \gamma (J_{ik}{-}J_{jk}))\Big].
\end{split}
\end{aligned}
\right.
\end{align}

\section{Estimation of Optimal Energies}
Here we will describe an estimate of the maximal value of a polynomial energy function in $n$ spins. 
In general, for vertices $V=\{1,\dots,n\}$, consider the cost function $C:\{-1,+1\}^V\rightarrow \RR$ defined by the polynomial 
\begin{align}
    C(z_1,\dots,z_n) & = \sum_{S\subseteq V}c_S z_S, 
\end{align}
where $z_S$ is shorthand for the product $\sum_{i\in S}z_i$ and the polynomial has $2^n$ coefficients $c_S\in \RR$. 
Without loss of generality we assume that $c_\varnothing = 0$, i.e.\ the function has zero offset. 
For a uniform $2^{-n}$ distribution over the possible strings $z$, the expected function value will be $0$, and the expected variance will be the sum of squares of its coefficients, i.e.\
\begin{align}
    \langle C \rangle  = \EE[C(Z)] & = 2^{-n}\sum_{z\in\{-1,+1\}^n}C(z) = 0 \\
    \langle C^2\rangle  = \EE[C^2(Z)] & = 2^{-n}\sum_{z\in\{-1,+1\}^n}C^2(z) = \sum_{S\subseteq V}c_S^2.
\end{align}
This is an exact result that holds for all polynomials $C$, and hence also for quadratic Ising problems.
The maximum that we are interested in is defined by
\begin{align}
M & = \max_{z\in \{-1,+1\}^n}C(z). 
\end{align}
The general idea that we use here is that $M$ can be approximated by the expected maximum among $2^n$ independent samples of a distribution that mimics picking a random string $z\in\{-1,+1\}^n$ and then returning the value $C(z)$. 
More precisely let $Z$ be the random variable that produces strings from $\{-1,+1\}^n$ with uniform probability $2^{-n}$,  then we want to estimate $M$ using the expectation 
\begin{align}
\hat M & = \EE\big[\max (C(Z_1),\dots,C(Z_{2^n}))\big].
\end{align}
Next we will argue that the distribution of the random variable $C$ will approximate a normal distribution, and that we can estimate $M$ by the expected maximum of $2^n$ samples of this normal distribution. 

Assume that we can model the random variable $C(Z)$ by the summation $\sum_{S\subseteq V} c_S R_S$ with $R_S$ a uniformly random spin value $\{-1,+1\}$.
This model is not exact as it ignores the correlations between terms like $z_{\{i,j\}}$ and $z_{\{j,k\}}$ and $z_{\{i,k\}}$, but it can nevertheless give us an intuition about the values of $C$. 
The expectation of $c_S R_S$ is $0$ and its variance is $c_S^2$. 
Hence for large $n$ the central limit theorem tells us that this sum will approximate the normal distribution $N(0,\sigma^2)$ with $\sigma^2 = \sum_S c^2S$. 

What remains is to calculate what to expect for the maximum among $2^n$ samples from $N(0,\sigma^2)$. 
When sampling $k$ times the normal distribution $N(0,1)$, the expected maximum of these samples can be approximated by $\sqrt{2 \ln k}$.
In our case here we have $k=2^n$ and the scale has to be multiplied by the standard deviation $\sigma$. 
Hence we reached the estimate 
\begin{align}
\hat{M} & \approx \sigma\sqrt{2 n \ln 2} = \sqrt{2n\ln 2}\sqrt{\sum_{S\subseteq V}c_S^2}. 
\end{align}
For G-set instances with $n=|V|$ and $\sum_S c_S^2 = |E|$ we thus have
\begin{align} \label{eq:G-estimator}
\hat{M} & \approx \sqrt{2 |V|\ln 2}\sqrt{|E|} \approx 1.18 \sqrt{|V||E|}.
\end{align}
For I-set instances with $n=|V|$ and $\sum_S c_S^2 = |E|+|V|$ we have 
\begin{align} \label{eq:I-estimator}
\hat{M} & \approx \sqrt{2 |V|\ln 2}\sqrt{|E|+|V|} \approx 1.18 \sqrt{|V|(|E|+|V|)}.
\end{align}

We reiterate that the above is an informal argument and can be made more rigorous in different ways, depending on the characteristics of the problem instance. 
For example, in the case of quadratic cost functions with coefficients $c_S\in\{0,1\}$, in the limit of large $|V|$ and with $P^* \approx 0.76321$ the Parisi constant, we know through the work of \cite{montanari2019optimization} that the maximum $M$ can be estimated by 
\begin{align} \label{eq:Montanari-estimator}
\hat{M} & = 2 P^* \sqrt{|V| |E| (1-2 |E|/|V|^2)}\\
& \approx 1.52642 \sqrt{|V| |E|}, \text{ for sparse graphs with $2|E|/|V|^2 \approx 0$.}
\end{align}

For quadratic cost functions with coefficients $c_S\in\{-1,0,+1\}$ and with the $-1$ and $+1$ couplings evenly distributed, another estimator directly follows from \cite{parisi1979infinite}
\begin{align} \label{eq:Parisi-estimator}
\hat{M} & = P^*\sqrt{2|V||E|} \approx 1.07934 \sqrt{|V||E|}.
\end{align}
Note how the informal estimator of Equation \ref{eq:G-estimator} sits between the rigorous ones of Equations \ref{eq:Montanari-estimator} and \ref{eq:Parisi-estimator}. 
A third estimator would also be needed for the I-set of Table~\ref{table:G-set_ising_field}. 
In this article we sidestep this diversity of rigorous estimators by using the informal estimators of Equations \ref{eq:G-estimator} and \ref{eq:I-estimator}.

\end{document}